\documentclass[aps,prx,reprint,groupedaddress]{revtex4-1}

\usepackage{color}
\usepackage{float}
\usepackage{graphicx}
\usepackage{epstopdf}  
\usepackage{psfrag}
\usepackage{epsfig}
\usepackage{subfigure}
\usepackage{mathtools}

\begin{document}


\title{Optical Isolation based on Space-time Engineered\\ Asymmetric Photonic Bandgaps}


\author{Nima Chamanara}
\author{Sajjad Taravati}
\author{Zo\'{e}-Lise Deck-L\'{e}ger }
\author{Christophe Caloz}
\affiliation{Poly-Grames Research Center, \'{E}cole Polytechnique de Montr\'{e}al,\\ Montr\'{e}al, Qu\'{e}bec H3T 1J4, Canada.}


\date{\today}

\begin{abstract}
Nonreciprocal electromagnetic devices play a crucial role in modern microwave and optical technologies. Conventional methods for realizing such systems are incompatible with integrated circuits. With recent advances in integrated photonics, the need for efficient on-chip magnetless nonreciprocal devices has become more pressing than ever. This paper leverages space-time engineered asymmetric photonic bandgaps to generate optical isolation. It shows that a properly designed space-time modulated slab is highly reflective/transparent for opposite directions of propagation. The corresponding design is magnetless, accommodates low modulation frequencies, and can achieve very high isolation levels. An experimental proof of concept at microwave frequencies is provided.
\end{abstract}


\maketitle

\section{Introduction}

Electromagnetic nonreciprocity plays a crucial role in modern electronic and optical technologies. Historically, breaking Lorentz reciprocity has been most often relying on magnetically biased magnetoelectric~\cite{gurevich1996magnetization, saleh1991fundamentals} and magnetoplasmonic~\cite{ishimaru1991electromagnetic, hartstein1974observation, montoya2009surface, chamanara2013_opex_isol, chamanara2013_opex_coupler, chamanara2013_prb_grapheneTEplasmon} materials. However,  magnetic materials are incompatible with integrated circuit technology. Moreover, such magnet-based technologies are based on bulky and expensive magnets. With the emergence of integrated photonics~\cite{soref2006past,pavesi2004silicon, reed2004silicon, jalali2006silicon}, generating on-chip optical nonreciprocity has become of paramount importance, and novel -- magnetless -- nonreciprocal technologies have therefore become required.

Over the past few decades, extensive efforts have been devoted to produce magnetless nonreciprocity, in order to eliminate the aforementioned issues associated with magnets and magnetic materials. An approach consists in using unilateral components such as transistors, which break Lorentz reciprocity from their semiconductor junction bias. This technology has been used for several decades in microwave nonreciprocal components and, more recently, in nonreciprocal metamaterials. Nonreciprocal transistor-based circulators~\cite{tanaka1965active, smith1988gaas, ayasli1989field, kalialakis2000analysis}, nonreciprocal metamaterials based on transistor-loaded unit cells~\cite{kodera2011artificialFaraday, kodera2013magnetless, sounas2013electromagnetic, kodera2012switchable, Joannopoulos_PNAS_2012, Taravati_2016_NR_Nongyro}, and nonreciprocal components based on staggered switched delay lines~\cite{reiskarimian2016magnetic} belong to this category. However, despite being compatible with integrated circuit technology, these devices suffer from relatively poor power handling and noise figure~\cite{carchon2000power}. Moreover, their application at terahertz and optical frequencies is impeded by the frequency limitation of transistor technology.

The Lorentz reciprocity theorem does not apply to nonlinear materials. This fact has spurred considerable efforts to achieve magnetless nonreciprocity and nonreciprocal devices based on nonlinearity~\cite{peng2013nonreciprocal, fan2012allsilicondiode, soljavcic2003nonlinear, gallo2001allopticaldiode, mahmoud2015allpassive, tocci1995thin}. This approach leverages the spatial asymmetry in the electromagnetic field intensity of a spatially asymmetric nonlinear permittivity profile for producing nonreciprocity. If nonlinearity is introduced at locations where the forward and backward waves have a significant difference in their electromagnetic field intensity, the forward and backward waves see different nonlinear permittivity terms and the structure hence exhibits nonreciprocity. However, since nonlinear effects only get pronounced at high signal levels, nonlinear techniques provide nonreciprocity only over a restricted signal power range. It was shown that in the presence of high-level input signals in a nonlinear optical isolator, some low-level signals get reciprocally transmitted, so that the structure does not really operate as a nonreciprocal optical component~\cite{shi2015limitationsNL}.

Balanced loss-gain media, also known as PT-symmetric media~\cite{ruter2010observation, lin2011unidirectionalPT, al2013unidirectional}, have been reported to exhibit unidirectional properties~\cite{peng2013nonreciprocalPT, chang2014parity, he2011parity, feng2013experimental, makris2010pt}. However, the nonreciprocity of the corresponding devices~\cite{chang2014parity, al2013unidirectional} is due again to nonlinearity rather than being a consequence of PT symmetry. Linear PT media are constrained to be reciprocal according to Lorentz reciprocity theorem and can not produce optical isolation~\cite{kulishov2005nonreciprocal, poladian1996resonance}.

Space-time modulation is another approach to break Lorentz reciprocity~\cite{yu2009complete, sounas2013giant, estep2014magnetic, hafezi2012optomechanically, Taravati_TAP_2016}. This approach is particularly suited for producing nonreciprocity at optical frequencies where transistor technology is unavailable. There have been several proposals to achieve magnetless nonreciprocity leveraging space-time variation. The technique proposed in~\cite{yu2009complete} uses oblique space-time interband transitions between two different modes of an optical waveguide. However, generating efficient coupling between the two waveguide modes, which are generally orthogonal, requires complex asymmetric modulation schemes. The techniques proposed in~\cite{sounas2013giant, estep2014magnetic} is based on counter rotating resonant modes with slightly shifted resonance frequencies. However, although it can achieve nonreciprocity over a subwavelength footprint, this approach requires sophisticated synchronized optical sources~\cite{sounas2013giant, estep2014magnetic}.

This paper introduces a novel concept for realizing optical isolation: space-time engineered asymmetric photonic bandgaps. In this approach, space-time variation in the permittivity of a medium is used to generate photonic band structures that are asymmetrically aligned with respect to the direction of propagation. It is shown that, with proper excitation, such a system can operate as a nonreciprocal (or unidirectional) optical device, i.e. an isolator. The modulation is uniform in the cross section of the waveguide, as opposed to that in~\cite{yu2009complete}, which leads to a much simpler structure. In addition, the required modulation frequency is relatively low, and may thus be conveniently provided by acoustic waves. The proposed approach may find applications in various integrated magnetless nonreciprocal optical systems. An experimental proof-of-concept at microwave frequencies is presented.

%
%

\section{Principle of operation} \label{sec:principles}

\begin{figure}[ht!]
\subfigure{\label{fig:concept_symgap}
\includegraphics[width=0.85\columnwidth]{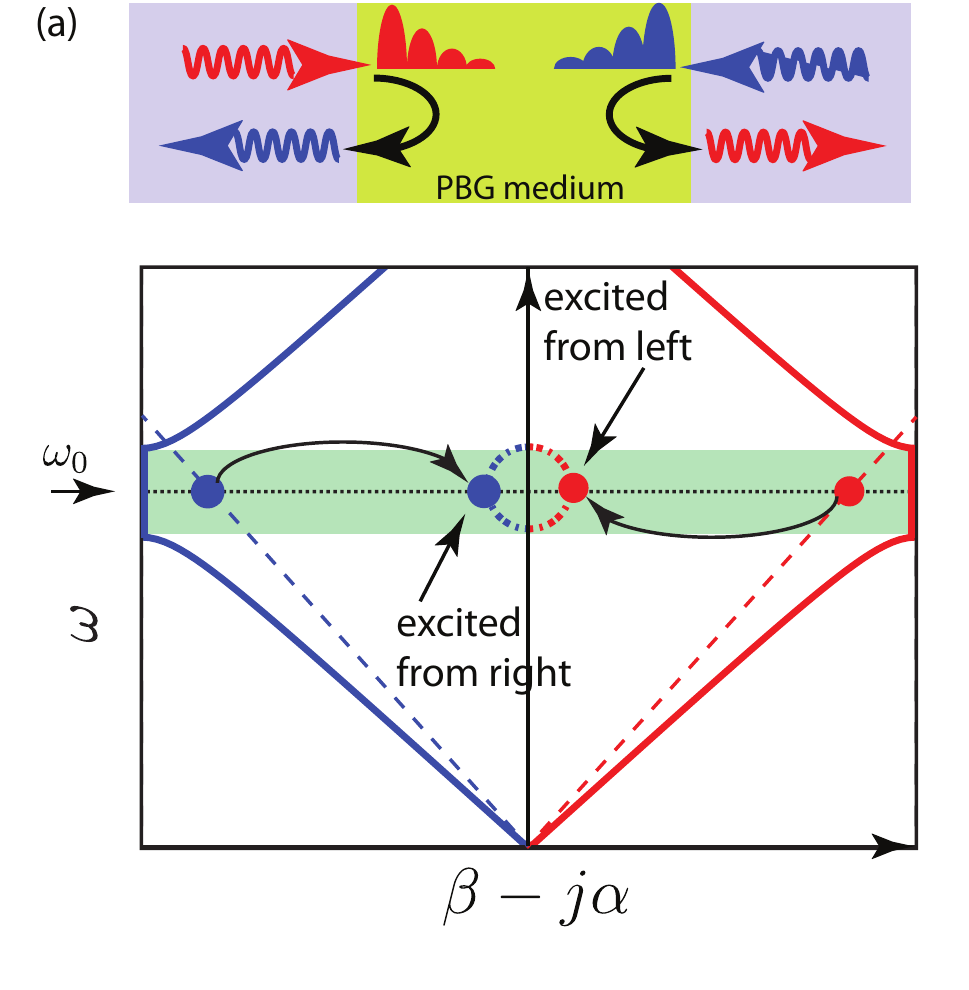}}
\subfigure{\label{fig:concept_asymgap}
\includegraphics[width=0.85\columnwidth]{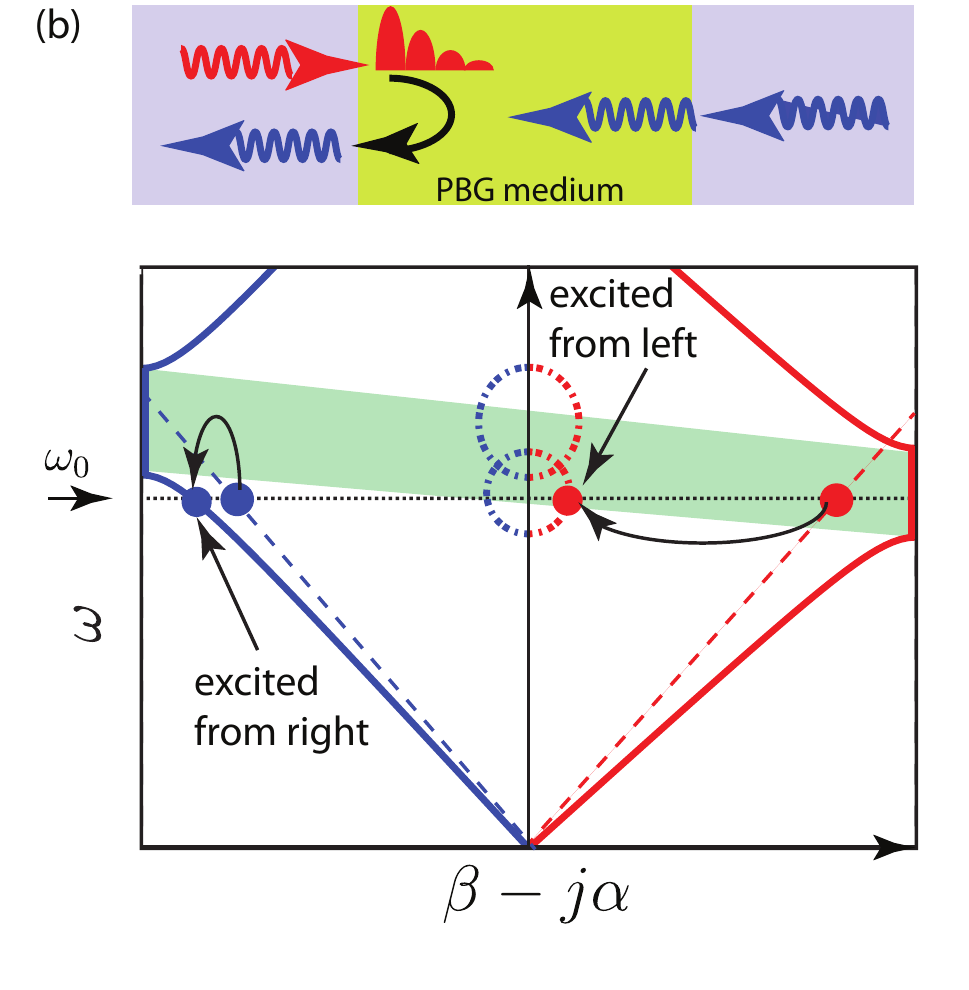}}
\caption{Principle of nonreciprocal Bragg reflection based on asymmetric photonic bandgaps. The red and blue colors represent forward and backward propagation, respectively. The dashed curves correspond to the dispersion curves of the input and output medium while the solid and dotted curves represent the real ($\beta$) and imaginary ($\alpha$) parts of the wavenumber, respectively, of the (central) photonic bandgap medium, assuming the harmonic time dependence $e^{j\omega t}$. The horizontal line corresponds to the excitation frequency, $\omega_0$. (a)~In a reciprocal system, the bandgap is symmetric with respect to positive and negative directions. Red/blue dots correspond to evanescent waves for excitation from the left/right. (b)~In a nonreciprocal system, the bandgap is tilted with a given slope. Red/blue dots correspond to evanescent and propagating waves, respectively, for excitation from the left/right. In one direction, the wave is totally reflected; in the opposite direction, it is fully transmitted.}
\label{fig:concept_asymgap_isolation}
\end{figure}

Consider a conventional reciprocal structure, such as for instance a Bragg grating or a waveguide filter, that supports photonic bandgaps, as illustrated in Fig.~\ref{fig:concept_symgap}. As the structure is composed of reciprocal materials, the bandgaps are perfectly horizontal in the dispersion diagram, i.e. symmetric with respect to positive and negative Bloch-Floquet wavenumbers. In the bandgaps, the Bloch-Floquet harmonics acquire an imaginary part in their wavenumber and hence become evanescent. Thus, when a wave incident on the structure is modulated at a frequency falling within a gap, it excites a complex, and hence evanescent, gap mode. This mode, marked by a red dot in Fig.~\ref{fig:concept_symgap}, decays exponentially. Therefore, assuming a proper choice of parameters, almost no power is transferred across the structure and, as a result of energy conservation, almost all of the incident power is reflected. Since the dispersion curves are symmetric with respect the wavenumber axis, when the structure is excited from the opposite end, the symmetric evanescent Bloch-Floquet mode, marked by the blue dot in Fig.~\ref{fig:concept_symgap}, is similarly excited, and most of the power is reflected.

Now consider a structure with an oblique, and hence asymmetric, bandgap, where the bandgap edges are different for the positive and negative directions, as shown in Fig.~\ref{fig:concept_asymgap}. When such a structure is excited from the left at the frequency corresponding to the horizontal line, the evanescent mode, marked by the red dot, is excited. If the structure is long enough, almost no power reaches the opposite end of it and the wave is fully reflected. In contrast, when the structure is excited from the right, the mode marked by the blue dot in Fig.~\ref{fig:concept_asymgap}, i.e. a propagating mode, is excited. Therefore, the incident electromagnetic power is transferred to the other side of the structure, and, assuming proper matching, is fully transmitted across it.

Producing such asymmetric dispersion curves requires a mechanism that breaks Lorentz reciprocity. In the next section, we use a space-time varying medium for that purpose. In such a medium, waves propagating in opposite directions perceive different dispersions, and the medium is therefore nonreciprocal. Corresponding dispersion curves are thus tilted with a given slope and hence form asymmetric bandgaps. We next analyze a finite space-time modulated slab and demonstrate its asymmetric-bandgap nonreciprocity.

\section{Unbounded space-time medium} \label{sec:infinite-space-time-media}

Consider an infinite space-time one-dimensionally periodic medium with permittivity
\begin{equation} \label{eq:general-moving-profile}
\epsilon(\mathbf{r}, t)=\epsilon_0\epsilon_r\left[1+Mf_\text{per}\left(t \pm z/v_\text{m}\right)\right],
\end{equation}
\noindent
where $f_\text{per}$ is a periodic function and $M$ is the modulation depth. This permittivity represents a periodic Bragg structure whose spatial profile moves in time at the modulation velocity $v_\text{m}$. Related space-time periodic media were first studied in the context of traveling-wave parametric amplification and parametric energy conversion~\cite{cullen1958travelling, tien1958parametric, oliner1959guided, cassedy1963dispersion, cassedy1967dispersion, chu1972wave, Taravati_TAP_2016}. The electric field in such a medium satisfies the following wave equation~\cite{cassedy1963dispersion}:
\begin{equation} \label{eq:wave-eq}
\nabla^{2}\mathbf{E}-\mu_0\frac{d^{2}}{dt^{2}}\left[\epsilon(\mathbf{r},t)\mathbf{E}\right]=0.
\end{equation}
\noindent
This equation admits solutions in the space-time Bloch-Floquet form~\cite{cassedy1963dispersion}
\begin{equation} \label{eq:Bloch-Floquet-form}
\mathbf{E}=e^{j\left(\omega t-\beta z\right)}\sum\limits_{n=-\infty}^{\infty}\mathbf{E}_{n}e^{jn\left(\omega_\text{m} t- k_\text{m} z\right)},
\end{equation}
\noindent
where $\omega_\text{m} = 2\pi/T$ is the modulation frequency, $T$ being the period of the function $f_\text{per}$, and $k_\text{m}=\omega_\text{m}/v_\text{m}$. The dispersion relation for the Bloch-Floquet waves in such a medium is found by substituting~\eqref{eq:Bloch-Floquet-form} and~\eqref{eq:general-moving-profile} into~\eqref{eq:wave-eq}. The resulting equation reduces, after truncation, to a matrix equation, $\mathbf{Ax}=\mathbf{0}$, through the orthogonality of the Fourier harmonics. Nullifying the determinant, i.e. setting $\left|\mathbf{A}\right|=0$, which is generally done numerically, provides the dispersion diagram of the medium.

\begin{figure}[ht!]
\includegraphics[width=0.9\columnwidth]{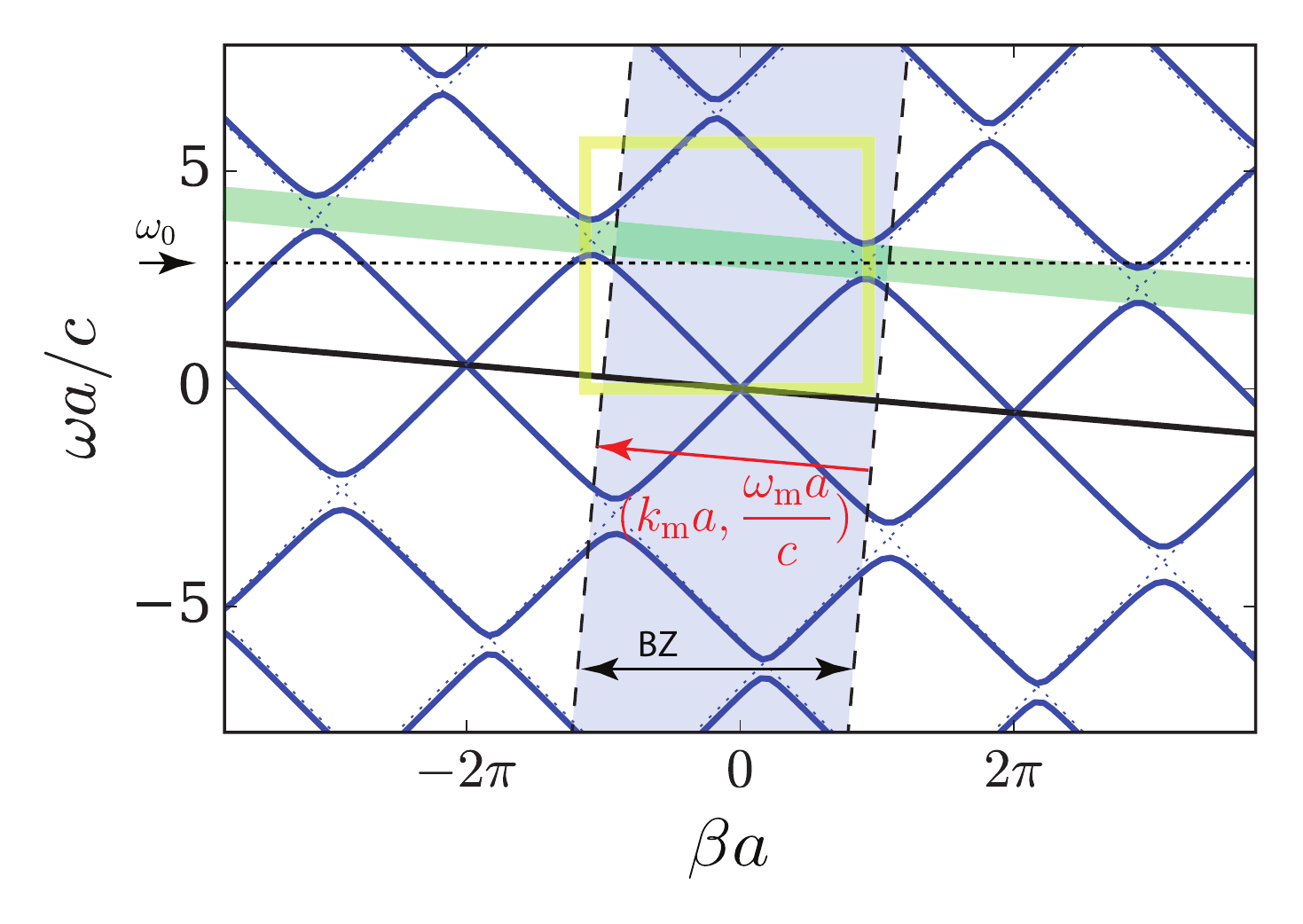}
\caption{Dispersion diagram for an infinite space-time modulated medium with permittivity given by~\eqref{eq:cos-profile} and parameters $\epsilon_r=12.25$, $M=0.5$, $\omega_\text{m}/\omega_0=0.13$ and $k_\text{m}/k_0=-2.27$. The dotted lines represent the limit of vanishingly small modulation depth $M\rightarrow0$. The red vector, $(k_\text{m} a, \omega_\text{m} a/c)$, corresponds to the space-time modulation wavenumber and frequency, where $a=2\pi/k_\text{m}$ is the spatial period and $c$ is the velocity of light. The dashed lines represent the Brillouin zone. The highlighted green region represents an oblique bandgap. The yellow window corresponds to the asymmetric bandgap structure in Fig.~\ref{fig:concept_asymgap}.}
\label{fig:st-inf-disp}
\end{figure}

In the following, we assume that the space-time profile or the periodic function, $f_\text{per}$, has a sinusoidal form. In this case, closed-form solutions can be derived for the eigenmodes and corresponding eigenvectors. An example of dispersion diagram for a space-time medium with permittivity
\begin{equation} \label{eq:cos-profile}
\epsilon(\mathbf{r}, t)=\epsilon_0\epsilon_r\left[1+M\cos\left(\omega_\text{m} t- k_\text{m} z\right)\right]
\end{equation}
is plotted in Fig.~\ref{fig:st-inf-disp}. It may be easily shown that each solution, $(\beta,\omega)$, corresponds to a mode formed by an infinite set of space-time harmonics, $(\beta\pm nk_\text{m},\omega\pm n\omega_\text{m})$ distributed along the vector $(k_\text{m}, \omega_\text{m})$. As a result, the Brillouin zone, represented by the dashed lines in Fig.~\ref{fig:st-inf-disp}, is tilted. In contrast to a purely spatial Bragg dispersion, the bandgaps appear asymmetric with respect to the positive ($\beta>0$) and negative $(\beta<0)$ directions of propagation, highlighted by the oblique green region in Fig.~\ref{fig:st-inf-disp}. The yellow window corresponds to the required asymmetric dispersion curve in Fig.~\ref{fig:concept_asymgap}. In the following section, this asymmetry is leveraged to generate space-time engineered optical isolation. However, accurate analysis of the structure requires taking into account all of the dispersion diagram, including the regions located outside the yellow window. We shall next develop an exact (full-wave) modeling technique to calculate the scattering parameters for a space-time modulated slab. The proposed method identifies all the modes excited inside the slab and provides physical insight into the scattering mechanism.

\section{space-time modulated slab} \label{sec:space-time-slab}

\begin{figure}[ht!]
\includegraphics[width=0.95\columnwidth]{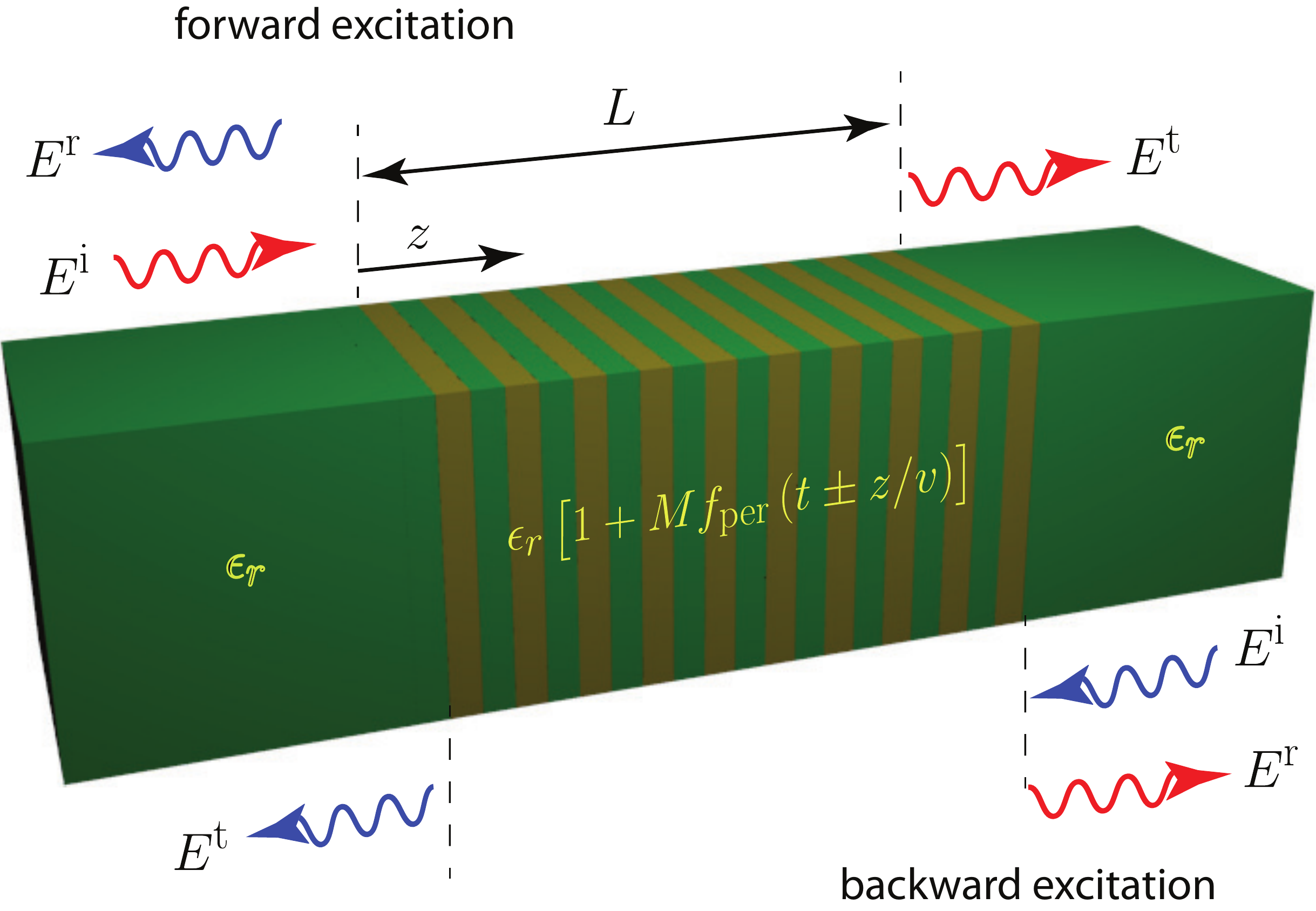}
\caption{Scattering from a space-time modulated slab. A finite part of a material with permittivity $\epsilon_r$ is spatio-temporally modulated with the space-time varying permittivity \eqref{eq:general-moving-profile} or~\eqref{eq:cos-profile}. The structure responds differently when excited from the left side or the right side, and is therefore nonreciprocal. Top arrows represent forward excitation, where the structure is excited from the left. Bottom arrows represent backward excitation, where the structure is excited from the right.}
\label{fig:st-slab}
\end{figure}

The asymmetric bandgaps in the dispersion diagram of a space-time-modulated slab may be leveraged for realizing optical isolation based on the principle explained in Sec.~\ref{sec:principles}. Consider the periodic space-time modulation~\eqref{eq:general-moving-profile} or~\eqref{eq:cos-profile} existing over a finite section of a background medium with permittivity~$\epsilon_r$, as shown in Fig.~\ref{fig:st-slab}. This structure may be analyzed with full-wave simulation techniques that can handle space-time varying media, such as the finite difference time domain (FDTD) method \cite{taflove2000computational}. However, such an analysis does not provide much insight into the operation mechanism. For gaining such insight, we shall use the mode-matching analysis technique.

In this technique, the structural modes and space-time Bloch-Floquet harmonics, excited inside the slab, are clearly identified. The electromagnetic fields in the incidence region, in the region at the other side of the slab, and the forward and backward propagating fields inside the slab are represented as superpositions of all the possible modal solutions in each region with unknown coefficients, corresponding to the weighting factors of the different modes. Details are provided in the next section.

\subsection{Mode-matching analysis}

\begin{figure}[ht!]
\subfigure{\label{fig:st-inf-disp-modes}
\includegraphics[width=0.95\columnwidth]{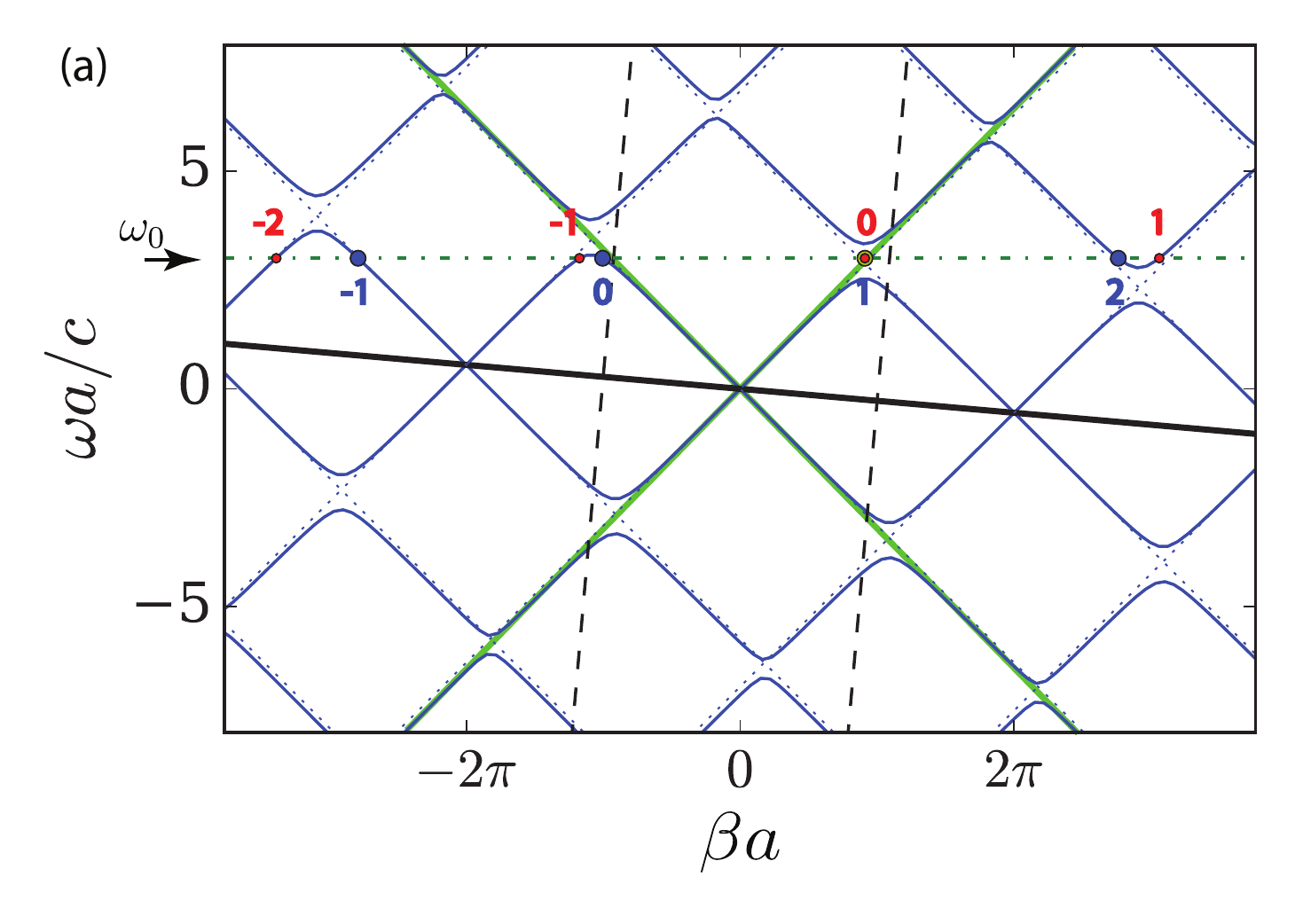}}
\hfill
\subfigure{\label{fig:st-inf-disp-modes-BZ}
\includegraphics[width=0.95\columnwidth]{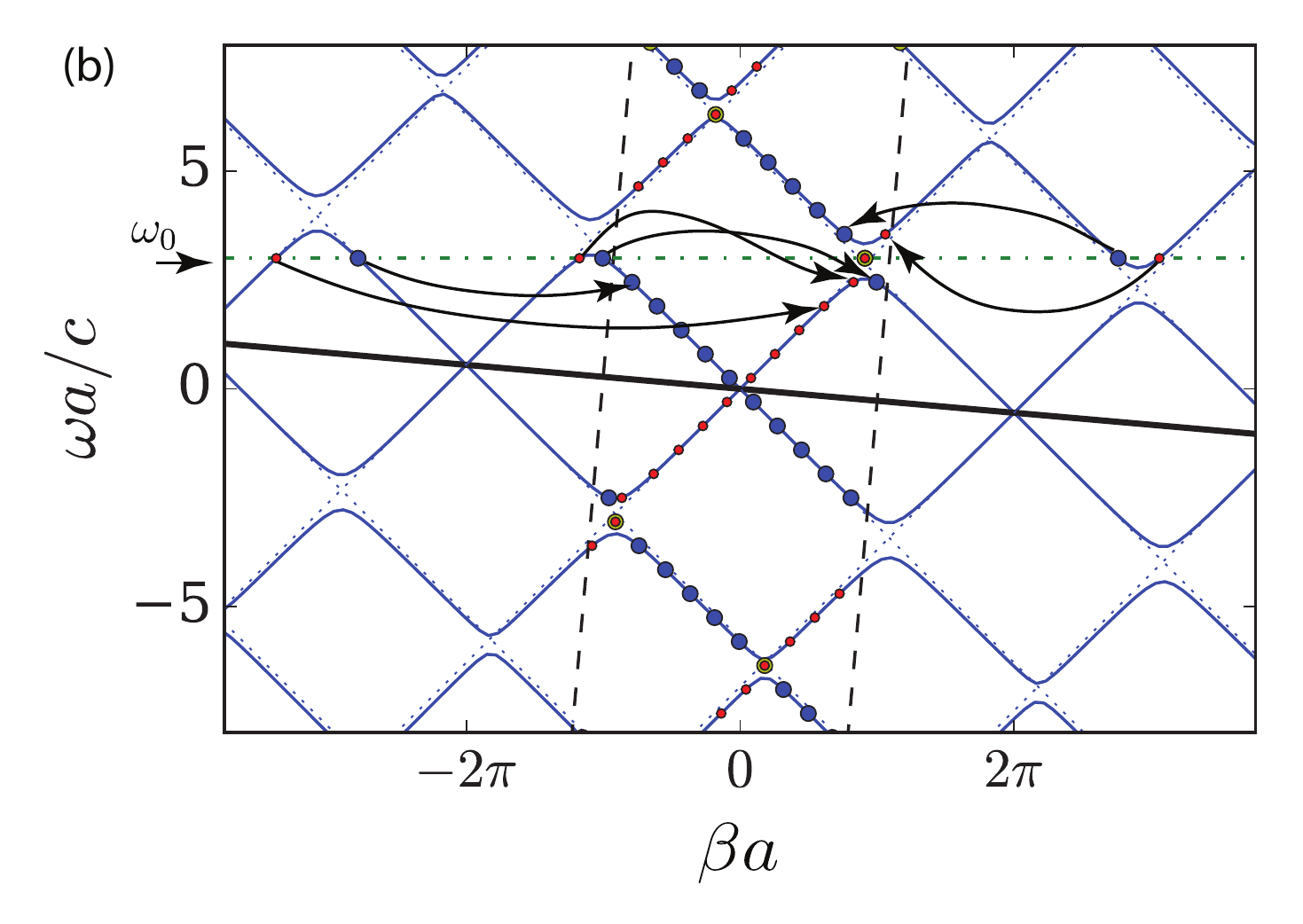}}
\caption{Bloch-Floquet modes excited in the space-time modulated slab of Fig.~\ref{fig:st-slab}, with the same parameters as in Fig.~\ref{fig:st-inf-disp}, by an incident wave with frequency $\omega_0$. The horizontal dashed line represents $\omega_0$ and the dots represent the corresponding excited modes. The red dots represent the modes with a positive group velocity while the blue dots represent the modes with a negative group velocity. The dashed lines delimit by the (oblique) Brillouin zone. The green lines represent the dispersion curves of the incident medium. (a)~General representation. (b)~Modes or space-time harmonics transferred into the Brillouin zone.}
\label{fig:st-inf-disp-modes-and-BZ}
\end{figure}

Consider a plane wave $\mathbf{E}^\text{i}=\hat{\mathbf{x}}e^{j(\omega t - kz)}$ incident on the space-time modulated slab sandwitched between media with permittivity $\epsilon_r$, as shown in Fig.~\ref{fig:st-slab}, where \mbox{$k = \omega \sqrt{\epsilon_r}/c$}. This wave will excite an infinite number of modes inside the slab so as to satisfy the boundary conditions on the two discontinuities delimiting the slab, and each of them will be formed by an infinite number of space-time harmonics. These modes are plotted in Fig.~\ref{fig:st-inf-disp-modes}, with red/blue dots corresponding to a given excitation frequency. We now decompose the total field of the forward problem (excitation from the left) into modes with positive group velocities, represented by the red dots,
\begin{equation} \label{eq:slab-modes-superposition-p}
\mathbf{E}^+(z,t) = \sum\limits_{p=-\infty}^\infty a_p^+ \mathbf{E}_p^+(z,t),
\end{equation}
\noindent
and the total field of the backward problem (excitation from the right) into modes with negative group velocities, represented by the blue dots,
\begin{equation} \label{eq:slab-modes-superposition-m}
\mathbf{E}^-(z,t) = \sum\limits_{p=-\infty}^\infty a_p^- \mathbf{E}_p^-(z,t).
\end{equation}

\noindent
In~\eqref{eq:slab-modes-superposition-p} and~\eqref{eq:slab-modes-superposition-m}, the terms $a_p^\pm$ represent the unknown modal coefficients, and each mode $p$ is represented as the space-time Bloch-Floquet expansion
\begin{equation}\label{eq:slab-modes-electric-pm}
\mathbf{E}_p^\pm(z,t)=\hat{\mathbf{x}}e^{j\left(\omega t-\beta_p^\pm z\right)}\sum\limits_{n=-\infty}^{\infty}E_{p,n}^\pm e^{jn\left(\omega_\text{m} t- k_\text{m} z\right)},
\end{equation}
\noindent
where $\beta_p$ represents the modal wavenumber, i.e. the projection of the dots onto the wavenumber (horizontal) axis. Our convention for numbering positive (red) and negative (blue) propagating modes is apparent in Fig.~\ref{fig:st-inf-disp-modes}, with the red/blue numbers corresponding to red/blue modes excited at the frequency $\omega_0$. Each of these numbers correspond to the index $p$ in~\eqref{eq:slab-modes-superposition-p} and \eqref{eq:slab-modes-superposition-m}. In a space-time modulated medium, all the modes excited at $\omega_0$ are distinct, as may be verified by transfer into the (oblique) Brillouin zone as shown in Fig.~\ref{fig:st-inf-disp-modes-BZ}. This transfer is achieved by shifting the modes outside the Brillouin zone in Fig.~\ref{fig:st-inf-disp-modes} by multiple integers of the oblique vector $(k_\text{m} a, \omega_\text{m} a/c)$ until they fall in the Brillouin zone. In a conventional static (or purely spatially modulated) Bragg structure, all the red/blue points would fold back onto the same red/blue point in the Brillouin zone, i.e. represent identical (linearly dependent) modes, so that all but one mode may be discarded. In contrast, in a space-time modulated medium, the modes numbered in Fig.~\ref{fig:st-inf-disp-modes}, are distinct (linearly independent), corresponding to different frequencies, and must all be taken into account for a complete description of the physics.

Consider first the forward problem (excitation from the left). The waves reflected and transmitted by the slab may be represented as superpositions of plane waves in the uniform medium with relative permittivity $\epsilon_r$, propagating in the $-z$ and $+z$ directions, respectively. In order to satisfy the boundary conditions, these waves must include all the temporal frequencies generated inside the slab, leading to the expansions
\begin{equation} \label{eq:refl-pw-sum}
\mathbf{E}^\text{r}(z,t)=\hat{\mathbf{x}}\sum\limits_{p=-\infty}^\infty a_p^\text{r} e^{j(\omega_p t + k_p z)},
\end{equation}
\begin{equation} \label{eq:trans-pw-sum}
\mathbf{E}^\text{t}(z,t)=\hat{\mathbf{x}}\sum\limits_{p=-\infty}^\infty a_p^\text{t} e^{j(\omega_p t - k_p z)},
\end{equation}
\noindent
where $\omega_p = \omega + p\omega_\text{m}$, $k_p = \omega_p\sqrt{\epsilon_r}/c$ and $a_p^\text{r}$, $a_p^\text{t}$ are unknown coefficients.

The magnetic field corresponding to each excited slab mode, namely $\mathbf{E}_p^\pm$ in \eqref{eq:slab-modes-electric-pm}, follows from the Maxwell-Faraday equation, \mbox{$\nabla \times \mathbf{E} = -\mu_0\frac{\partial }{\partial t}\mathbf{H}$}, leading to the modal expansion
\begin{equation}
\mathbf{H}_p^\pm(z,t)=\hat{\mathbf{y}}e^{j\left(\omega t-\beta_p^\pm z\right)}\sum\limits_{n=-\infty}^{\infty}H_{p,n}^\pm e^{jn\left(\omega_\text{m} t- k_\text{m} z\right)},
\end{equation}
\noindent
where
\begin{equation}
H_{p,n}^\pm = \frac{\beta_p^\pm+n k_\text{m}}{\mu_0(\omega + n\omega_\text{m})} E_{p,n}^\pm.
\end{equation}
\noindent
Application of the boundary conditions, i.e. continuity of the tangential electric and magnetic fields at the slab interfaces, leads then to a system of equations for the unknown coefficients, whose solutions provide the reflected and transmitted fields as well as the fields inside the slab.
%
%
%
%
%
%

\subsection{Modal Distribution and Frequency Transitions} \label{sec:results}

Consider a space-time modulated slab with dispersion curves shown in Fig.~\ref{fig:st-inf-disp-modes}, where the modulation frequency is tuned such that the incident frequency excites the gap mode (red index 0) in the forward direction and a propagating mode (blue index 0) in the backward direction, as shown in Fig.~\ref{fig:st-inf-disp-modes}. As explained in the previous section, due to the tilt of the space-time diagrams, an infinite number of slab modes are excited. The operation of the device depends on the relative excitation strength of these modes. When the structure is excited from the left/right, these modes are excited with different weighting factors, i.e. the structure is nonreciprocal. This section quantifies the reflection and transmission, and corresponding isolation, as well as the modes inside the slab, for excitation the from the left/right.


\begin{figure}[ht!]
\subfigure{\label{fig:st-slab-RT-modes-FWD}
\includegraphics[width=0.95\columnwidth]{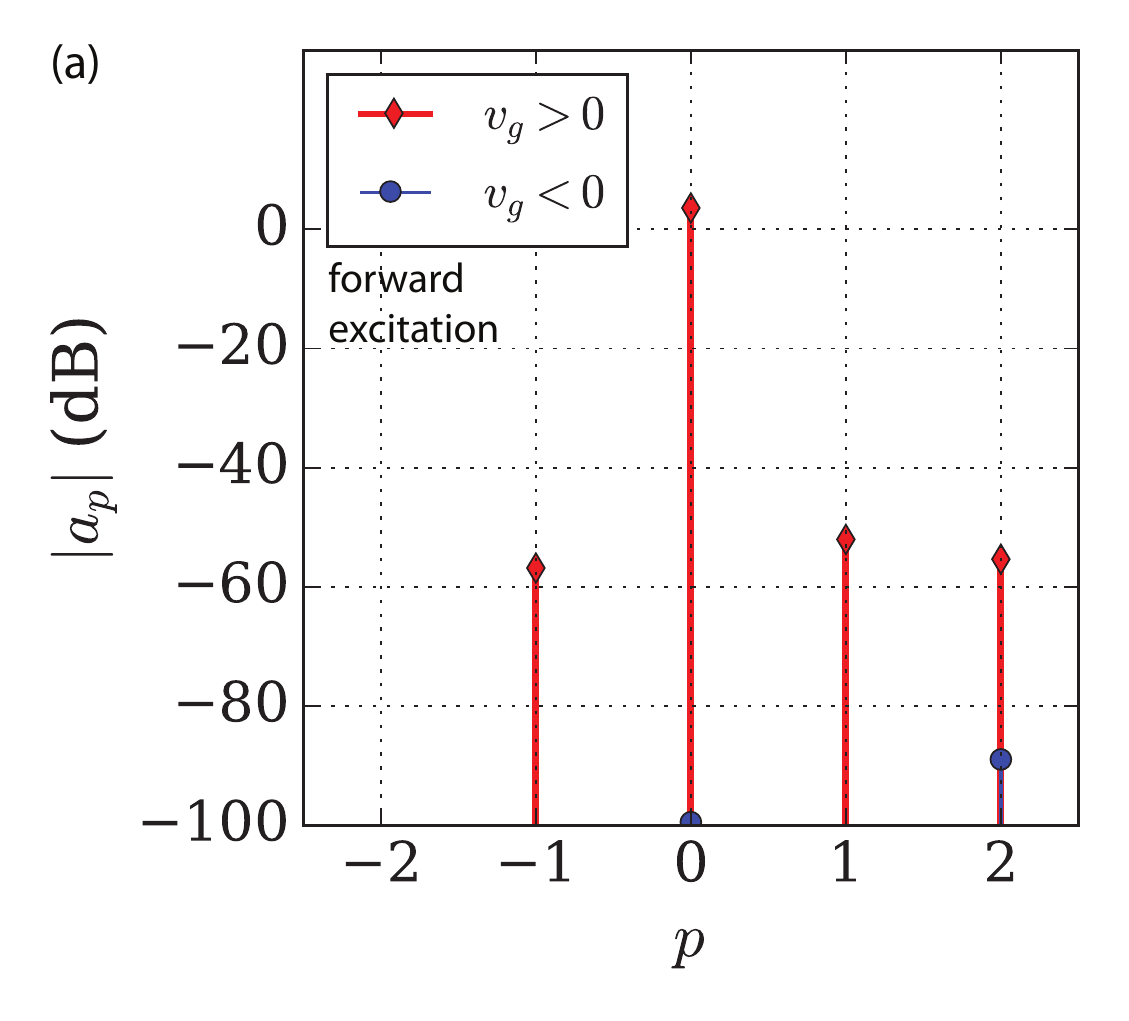}}
\subfigure{\label{fig:st-slab-RT-modes-BWD}
\includegraphics[width=0.95\columnwidth]{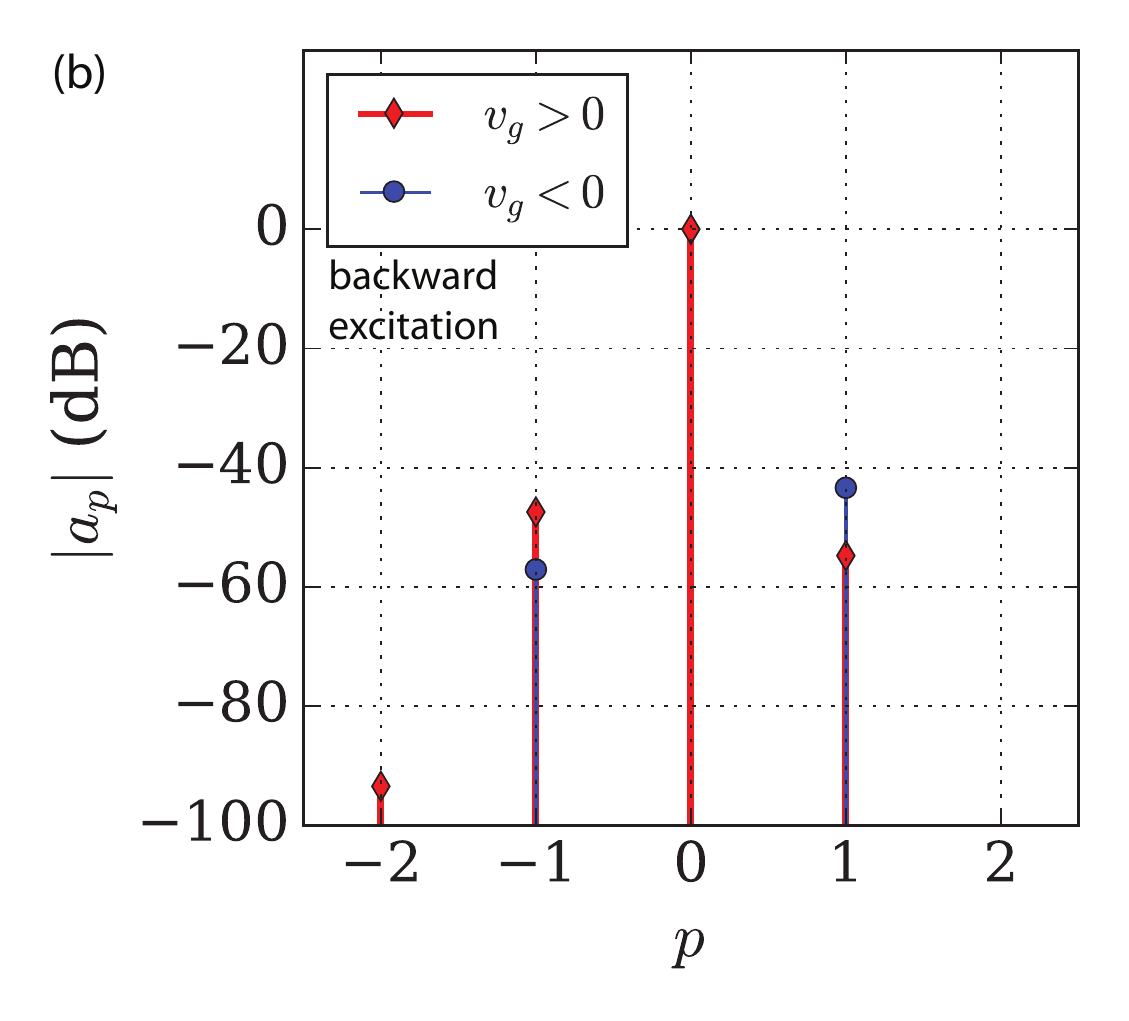}}
\caption{Magnitude of the modes excited in the space-time modulated slab of Fig.~\ref{fig:st-inf-disp-modes-and-BZ} with $L=200\lambda$. The normalized excitation frequency is $\omega_0=0.259c/a$. (a)~Slab excited from the left. The gap mode (red index~0 in Fig.~\ref{fig:st-inf-disp-modes}) is excited dominantly. All the other forward propagating modes (red) and backward propagating modes (blue) are very weakly excited. (b)~Slab excited from the right. The dominantly excited mode is a propagating mode (blue index~0 in Fig.~\ref{fig:st-inf-disp-modes}).}
\label{fig:st_inf_disp}
\end{figure}

\subsubsection{Excitation from the left} \label{sec:excitation-left}
Consider a space-time modulated slab with space-time permittivity \eqref{eq:cos-profile}, background permittivity $\epsilon_r=12.25$, modulation depth $M=0.02$, temporal and spatial modulation frequencies $\omega_\text{m}=0.13\omega_0$ and $k_\text{m}=-2.27k_0$, respectively, and length $L=200\lambda$, excited at the normalized frequency $\omega_0=0.259c/a$, where $a$ is the spatial period of the space-time modulated slab. The corresponding permittivity profile represents a sinusoidal Bragg grating, whose permittivity perturbation propagates towards the left inside the space-time modulated region, with velocity $v_\text{m}=-|\omega_\text{m}/k_\text{m}|$.

For excitation from the left, the amplitude of the modes excited inside the slab, calculated by the mode-matching analysis presented in the previous section, are presented in Fig.~\ref{fig:st-slab-RT-modes-FWD}. The red lines/diamonds correspond to positive group velocity (forward propagating) Bloch-Floquet modes, while the blue lines/circles correspond to negative group velocity (backward propagating) modes. It appears that mode $p=0$ is much more excited, by at least 50~dB, than the other ones, which indicates that all the modes falling outside the highlighted yellow window in Fig.~\ref{fig:st-inf-disp-modes} play an insignificant role, and that the performance of the space-time modulated slab can be closely predicted by the intuitive picture presented in Fig.~\ref{fig:concept_asymgap}. The reason why the mode $p=0$ is so much more excited than the others is because it is the only one that is close to the incident medium dispersion curve, as seen in Fig.~\ref{fig:st-inf-disp-modes}, and hence the only well phase- and impedance-matched to the incident medium.

\begin{figure}[ht!]
\subfigure{\label{fig:st-slab-RT-FWD}
\includegraphics[width=0.95\columnwidth]{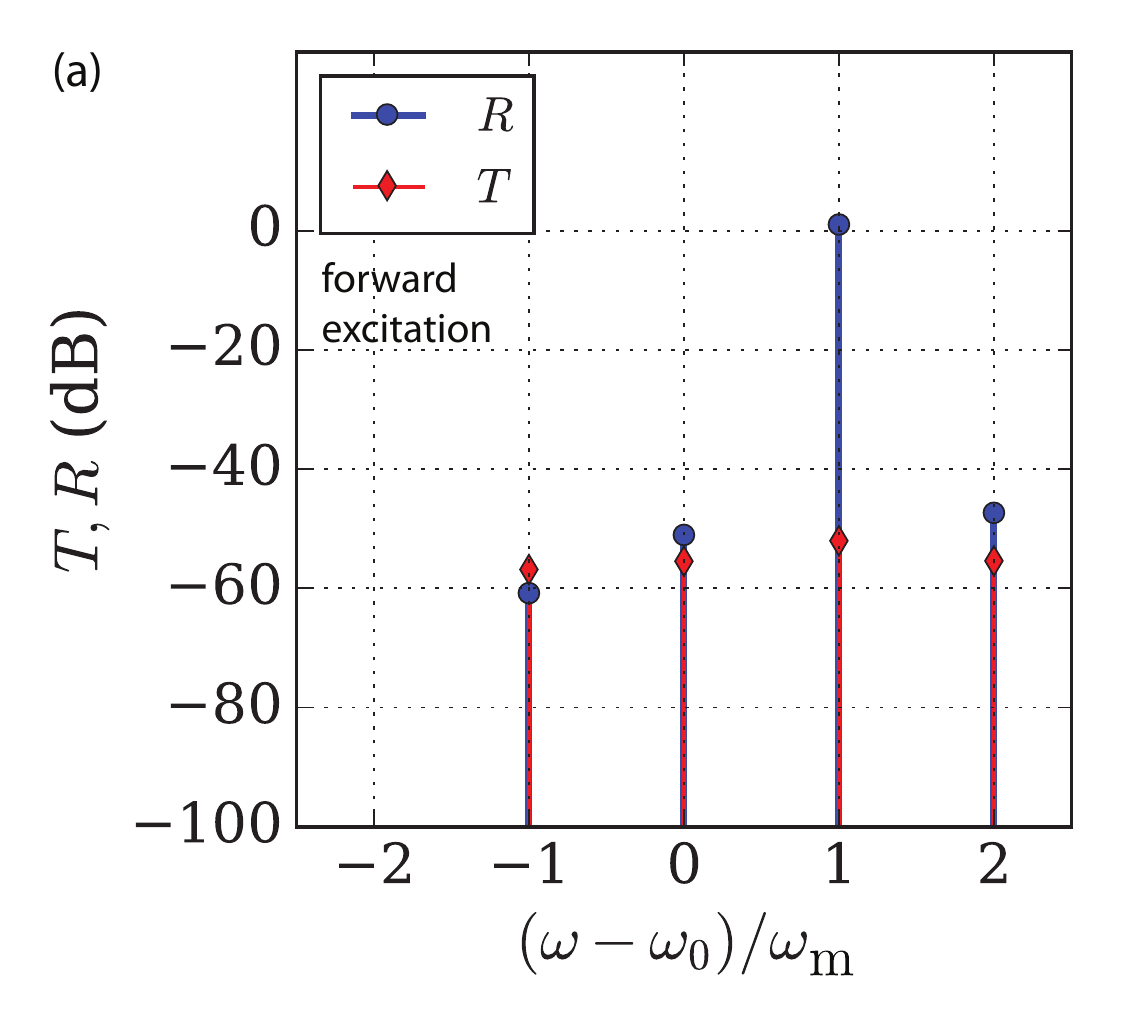}}
\subfigure{\label{fig:st-slab-RT-BWD}
\includegraphics[width=0.95\columnwidth]{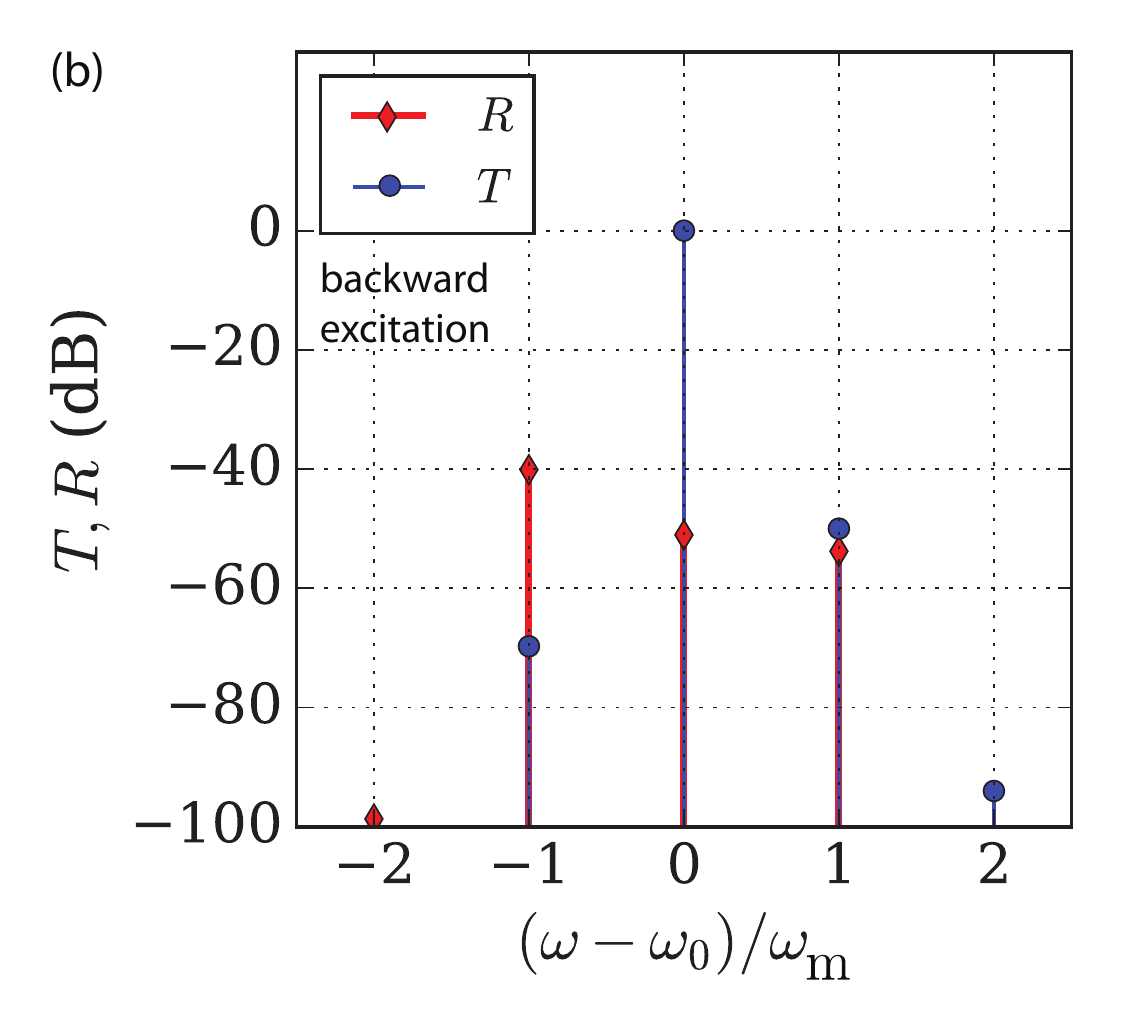}}
\caption{Reflection and transmission (outside the modulated slab) for the space-time modulated slab with the same parameters as in Fig.~\ref{fig:st_inf_disp}. (a)~Slab excited from the left. The dominantly excited evanescent gap mode decays exponentially and conveys no power to the transmitted region. All of the power is reflected. The reflected wave is blue-shifted. (b)~Slab excited from the right. The dominantly excited propagating mode transfers all its energy to the other end. Almost all the power is transmitted at the fundamental frequency ($\omega_0$).}
\label{fig:st_inf_TR}
\end{figure}

Moreover, Fig.~\ref{fig:st-inf-disp-modes} shows that this mode falls in a bandgap of the modulated structure. It is thus evanescent and exponentially decaying in the modulated structure, carrying almost no power to its right end. Since the system is assumed to be lossless, the incident power can only be reflected towards to input medium. This is confirmed in Fig.~\ref{fig:st-slab-RT-FWD}, which plots the transmitted and reflected amplitudes for different temporal frequency harmonics. The transmission level is below $-40$~dB for all frequency harmonics, and the power is almost fully reflected at the blue-shifted frequency $\omega_0 + \omega_\text{m}$.  This is a space-time blue Doppler shift due to the fact that the space-time varying medium profile has an opposite (negative) phase velocity, $v_\text{m}=\omega_\text{m}/k_\text{m}$, with respect to the source on the left. This effect will be detailed in Sec.~\ref{sec:blueshift}.

The levels of the transmitted and reflected power may be controlled by tuning the modulation depth and the length of the slab. For a given modulation depth, it is always possible to reduce the transmitted power to a desired level by increasing the length of the slab. Notice that the reflected power is slightly greater than unity. This is not at odds with energy conservation since energy is pumped into the space-time varying medium.

\subsubsection{Excitation from the right} \label{sec:excitation-right}

For excitation from the right, the amplitudes of the modes excited in the slab are plotted in Fig.~\ref{fig:st-slab-RT-modes-BWD}. The negative group velocity (backward propagating) mode $p=0$ is excited much more, by at least 40~dB, than the others, because it is much better matched to the incident wave. As seen in Fig.~\ref{fig:st-inf-disp-modes}, this slab mode is a propagating one, and it therefore carries almost all the power to the other end. Hence, the structure is expected to be highly transparent. This is confirmed in Fig.~\ref{fig:st-slab-RT-BWD}. Almost all the power is transmitted at the incident frequency, and the reflected power from the slab is below $-40$~dB for all the harmonics. The amount of reflected power is proportional to the mismatch between the space-time modulated and incident media, which is in turn proportional to the modulation depth.

\subsubsection{Explanation of the Doppler shift in the reflected wave} \label{sec:blueshift}

Although not including any matter motion, the slab medium in Fig.~\ref{fig:st-slab} supports space-time \emph{perturbation motion}~\eqref{eq:cos-profile}. This is why the wave reflected from the bandgap structure experiences the temporal frequency shift observed in Fig.~\ref{fig:st-slab-RT-FWD}. We shall next show that this shift, from $\omega_0$ to $\omega_0+\omega_\text{m}$, and hence of magnitude $\omega_\text{m}$, corresponds to the conventional relativistic Doppler shift for a wave reflected from a moving medium with the velocity $v_\text{m}$,
\begin{equation} \label{eq:doppler-relativitic}
\Delta \omega = \left(\frac{1+|v_\text{m}|/c}{1-|v_\text{m}|/c}-1\right)\omega_0.
\end{equation}

\begin{figure}[ht!]
\includegraphics[width=0.95\columnwidth]{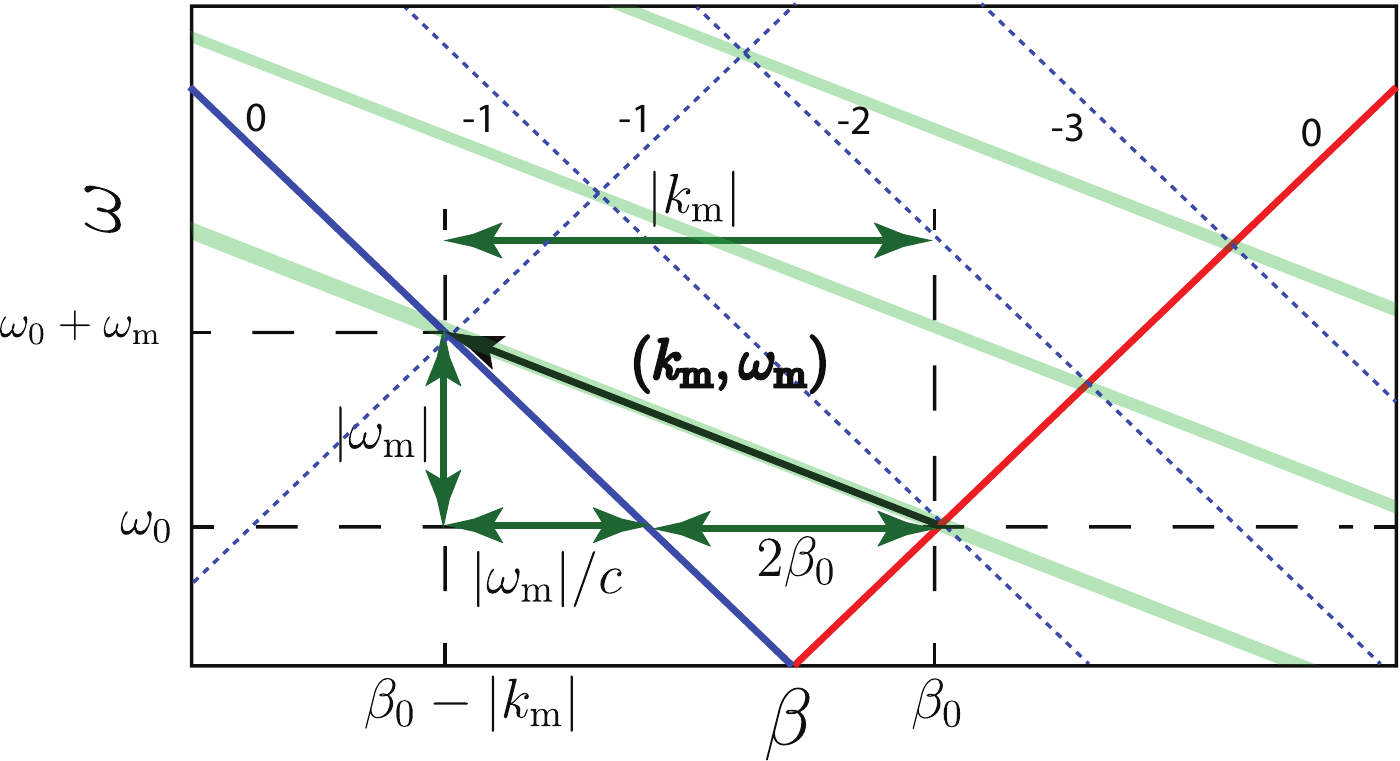}
\caption{Dispersion diagram and geometrical parameters in the limit $M\rightarrow 0$ for explaining the Doppler shift of the reflected wave in Fig.~\ref{fig:st-slab-RT-FWD}. The solid lines represent the dispersion curves of the background medium and $c$ represents the velocity of light in this medium. The green bands correspond to the infinitesimal gaps.}
\label{fig:doppler-shift-expl}
\end{figure}

Figure~\ref{fig:doppler-shift-expl} shows the dispersion diagram for an infinitesimal modulation depth and the corresponding geometrical parameters related to the frequency shift in Fig.~\ref{fig:st-slab-RT-FWD}. Note that at the bandgap corresponding to the spatial and temporal frequencies $(\beta_0, \omega_0)$, the forward harmonic $n=0$ crosses the backward harmonic harmonic $n=-1$. Since the backward harmonic $n=-1$ is a version of the backward harmonic $n=0$ that is shifted by the vector $-(k_m, \omega_m)$, the endpoint of the vector $(\beta_0, \omega_0) + (k_m, \omega_m)$ lies at the intersection of the backward dispersion curve $n=0$ and the forward dispersive curve $n=-1$, as shown in Fig.~\ref{fig:doppler-shift-expl}. This leads to the geometrical relation
\begin{equation} \label{eq:doppler-k-m}
|k_\text{m}|=2\beta_0+|\omega_\text{m}|/c
\end{equation}
%

\noindent highlighted in the figure. Therefore the velocity of the space-time medium reads

\begin{equation} \label{eq:v-m}
|v_\text{m}| = \frac{|\omega_\text{m}|}{|k_\text{m}|} = \frac{|\omega_\text{m}|}{2\beta_0+|\omega_\text{m}|/c}.
\end{equation}

\noindent
Substituting \eqref{eq:v-m} into \eqref{eq:doppler-relativitic} results in the observed frequency shift, $\Delta \omega=|\omega_\text{m}|$, which shows that perturbation motion leads to the same Doppler effect as matter motion. Note that this is true \emph{only in the absence of dispersion}, corresponding to the straight line condition of~\eqref{eq:v-m}, while introducing dispersion would allow one depart from~\eqref{eq:doppler-relativitic} and engineer the Doppler shift.

The frequency shift may also be explained in terms of intraband photonic transitions between the forward and backward propagating modes of a single-mode waveguide. For small modulation depth ($M \ll 1$), instead of considering the exact periodic problem involving the infinite set of space-time harmonics, the problem may be approximated as follows. As an electromagnetic wave with momentum and frequency $(k_0, \omega_0)$ in the background medium penetrates into the space-time modulated section, the space-time medium provides the extra momentum and energy corresponding to $\pm(k_\text{m}, \omega_\text{m})$ to the wave. If the resulting momentum and energy $(k_0\pm k_\text{m}, \omega_0\pm\omega_\text{m})$ correspond to a mode of the waveguide, coupling to this mode occurs and the incoming waveguide mode is then gradually transformed into the waveguide mode at $(k_0\pm k_\text{m}, \omega_0\pm\omega_\text{m})$. In contrast, if $(k_0\pm k_\text{m}, \omega_0\pm\omega_\text{m})$ does not correspond to a mode of the waveguide, the corresponding wave passes through the space-time modulated region almost unaffected. This interband transition picture and the associated coupled mode analysis are accurate only for very small modulation depths, and should therefore be considered with great care in the case of strong modulations, as it ignores the rich spectral features of the electromagnetic band structure of the space-time modulated system. Nonetheless, this explanation provides an alternative intuitive understanding of the Doppler frequency shift described above.

\begin{figure}[ht!]
\subfigure{\label{fig:interband_transition_2f}
\includegraphics[width=1.0\columnwidth]{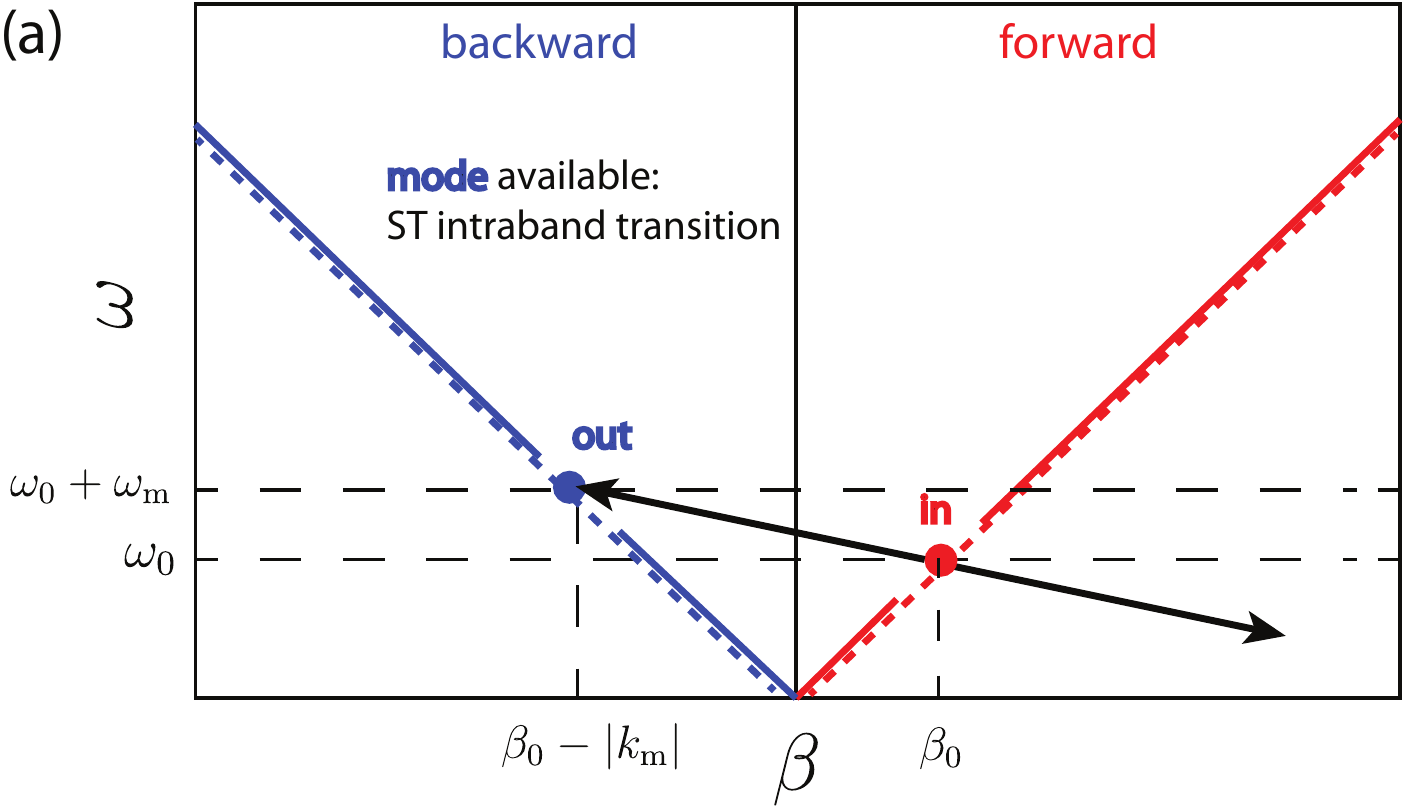}}
\subfigure{\label{fig:interband_transition_2b}
\includegraphics[width=1.0\columnwidth]{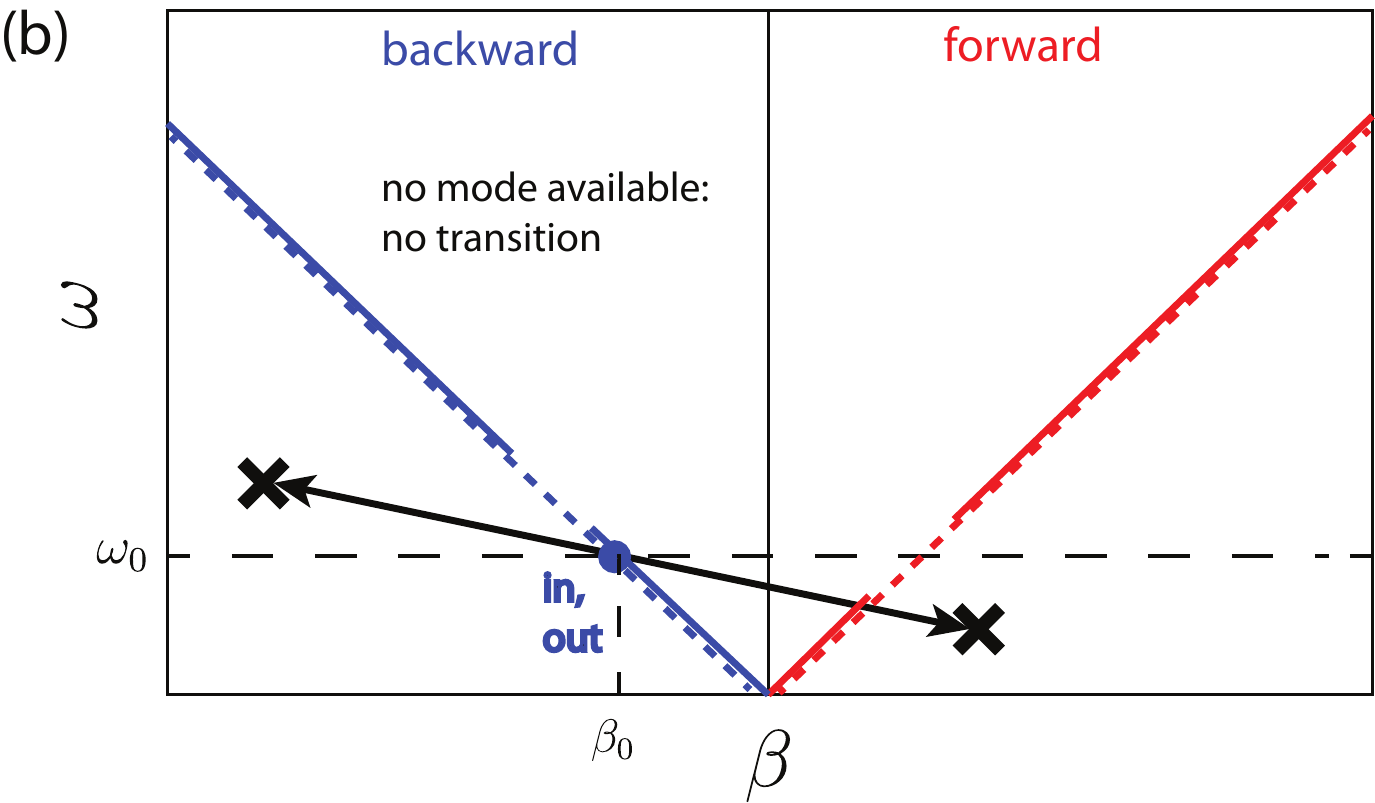}}
\caption{Explanation of the frequency up-shift in the reflected field in terms of intraband photonic transitions. The dashed lines correspond to the dispersion curves of the background medium. The arrows represent the momentum end energy provided by the space-time modulated region, $\pm(k_\text{m}, \omega_\text{m})$. (a)~Left excitation: the forward propagating mode gradually transforms into a backward propagating mode with frequency $\omega_0+\omega_\text{m}$ and is reflected. (b)~Right excitation: the propagating mode passes through the space-time section as $(k_0\pm k_\text{m}, \omega_0\pm\omega_\text{m})$ does not correspond to any waveguide mode.}
\label{fig:interband_transition_2}
\end{figure}

For the space-time modulated problem considered in the Sec.~\ref{sec:infinite-space-time-media}, the dispersion curves of the single mode background medium and the corresponding momentum and energy, $\pm(k_\text{m}, \omega_\text{m})$, provided by the space-time medium, are plotted in Fig.~\ref{fig:interband_transition_2f}, for excitation from the left. As $(k_0+ k_\text{m}, \omega_0+\omega_\text{m})$ corresponds to a backward propagating mode of the background medium, the incident forward propagating mode gradually transforms to a blue-shifted backward propagating mode, i.e. reflects with a frequency up-shift exactly equal to $\Delta\omega = \omega_\text{m}$. In contrast, for a wave exciting the space-time modulated region from the right, the corresponding momentum and energy, $\pm(k_\text{m}, \omega_\text{m})$ provided by the space-time medium, is plotted in Fig.~\ref{fig:interband_transition_2b}. As $(k_0+ k_\text{m}, \omega_0+\omega_\text{m})$ does not correspond to a mode of the background medium, it passes through the space-time region almost unaffected.

\begin{figure}[ht!]
\subfigure{\label{fig:E-profile-gapmode}
\includegraphics[width=0.95\columnwidth]{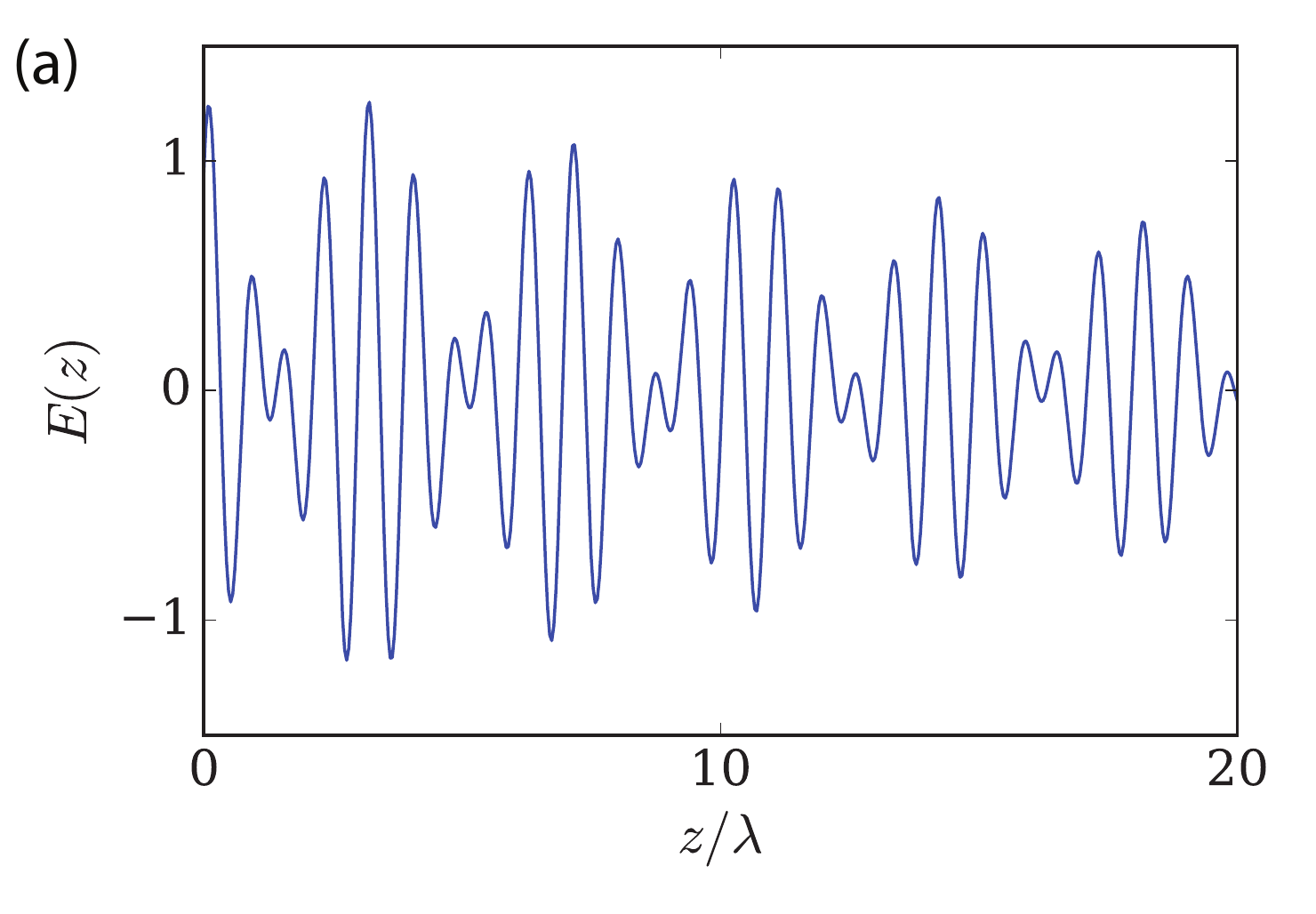}}
\subfigure{\label{fig:st-harmonics-gapmode}
\includegraphics[width=0.95\columnwidth]{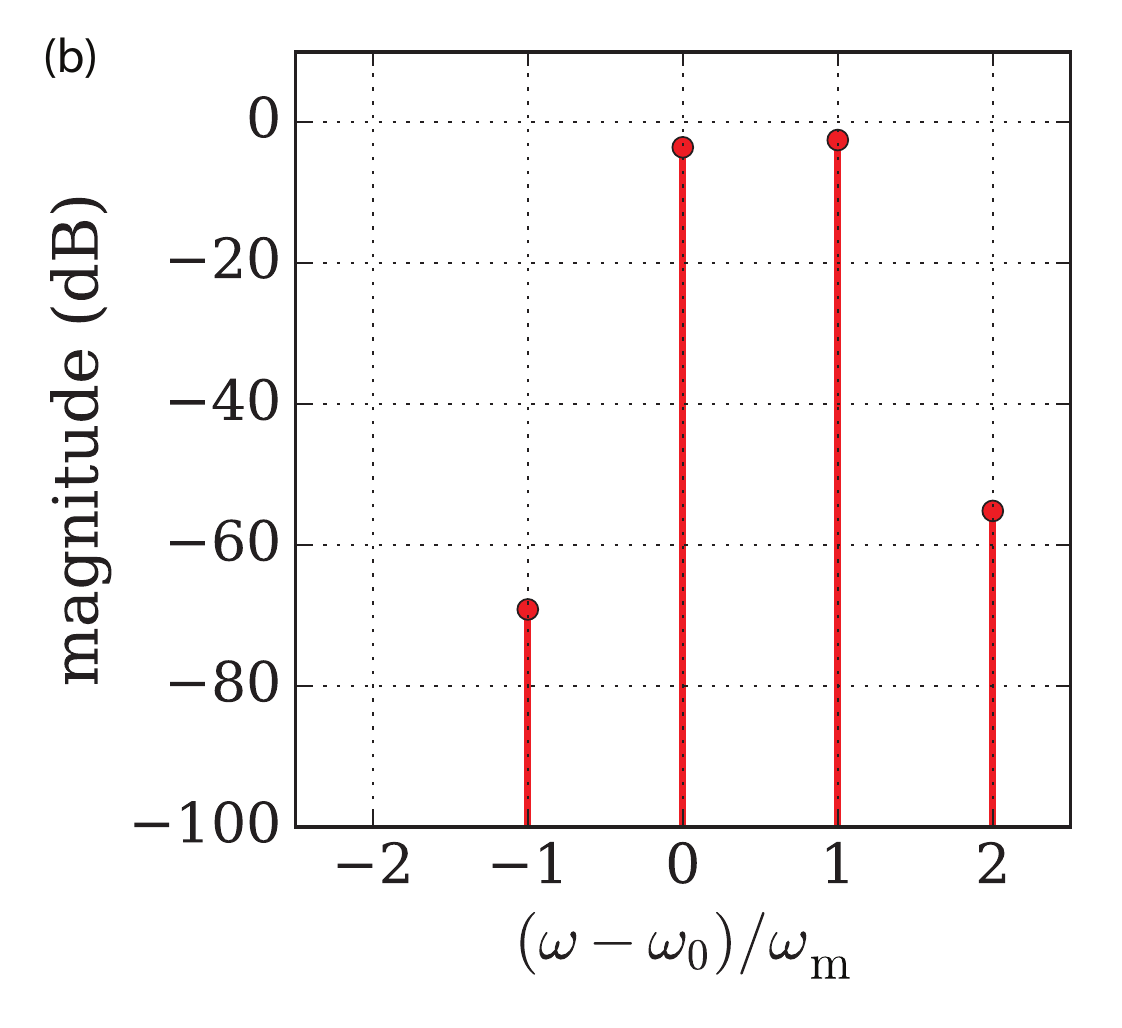}}
\caption{Electric field pattern and frequency spectrum of the gap mode for excitation from the left. (a)~Electric field pattern for the evanescent gap mode, corresponding to the red index~0 in Fig.~\ref{fig:st-inf-disp-modes}. (b)~Frequency harmonics of the evanescent gap mode plotted in (a). This mode has only two significant harmonics, at frequencies $\omega_0$ and $\omega_0+\omega_\text{m}$. }
\label{fig:gapmode}
\end{figure}

\begin{figure}[ht!]
\includegraphics[width=0.9\columnwidth]{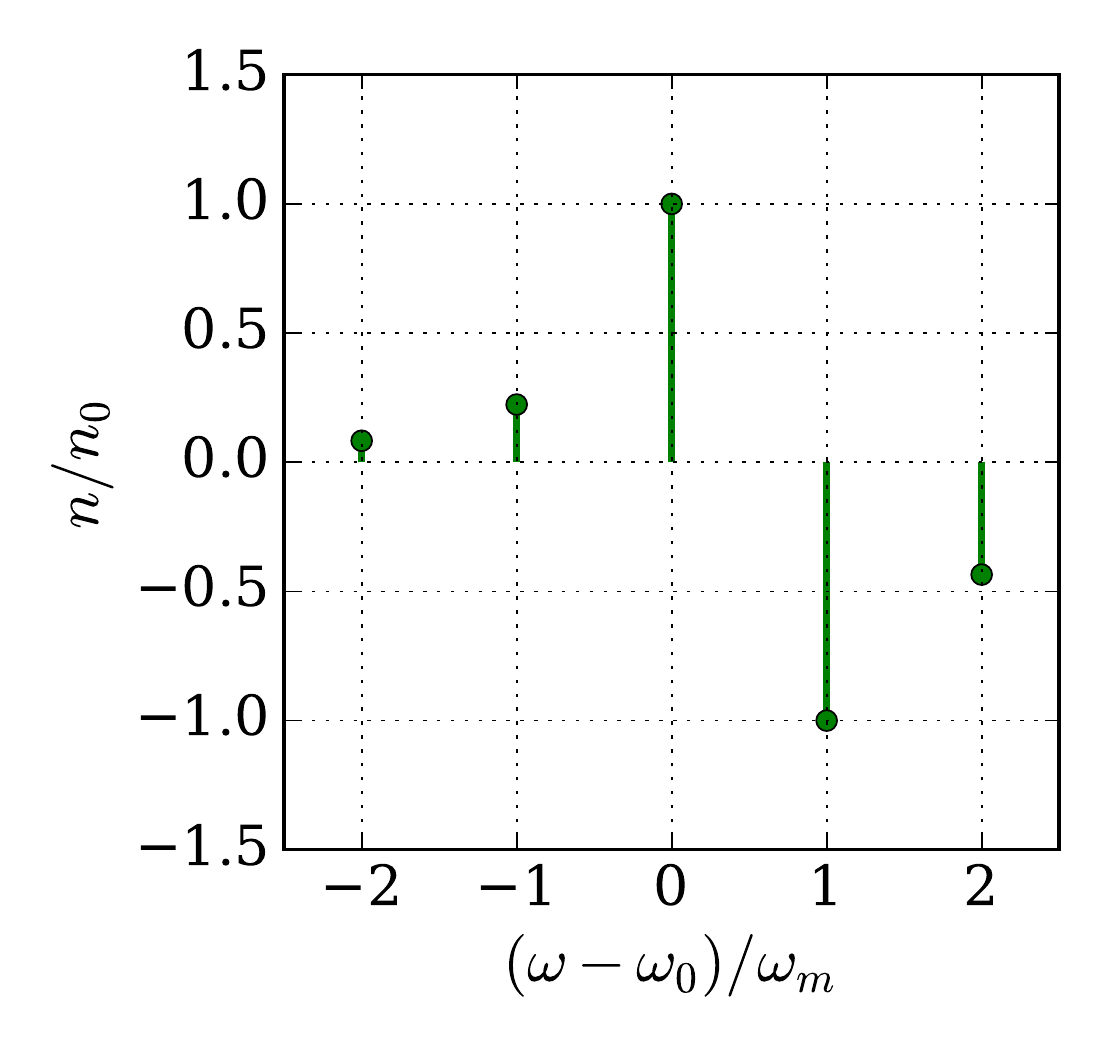}
\caption{Effective refractive index for the different harmonics of the gap mode plotted in Fig.~\ref{fig:gapmode}. Each point represents the effective refractive index corresponding to the ratio $\eta_0 H_n/E_n$ for the frequency harmonic $\omega_n=\omega_0+n\omega_\text{m}$. The harmonic at frequency $\omega_0$ has a positive refractive index (forward propagating), and is fully matched to the incident region. The harmonic at frequency $\omega_0+\omega_\text{m}$ has a negative refractive index (backward propagating), and is fully matched to the incident region.}
\label{fig:gapmode-impedance}
\end{figure}

\begin{figure}[ht!]
\includegraphics[width=0.9\columnwidth]{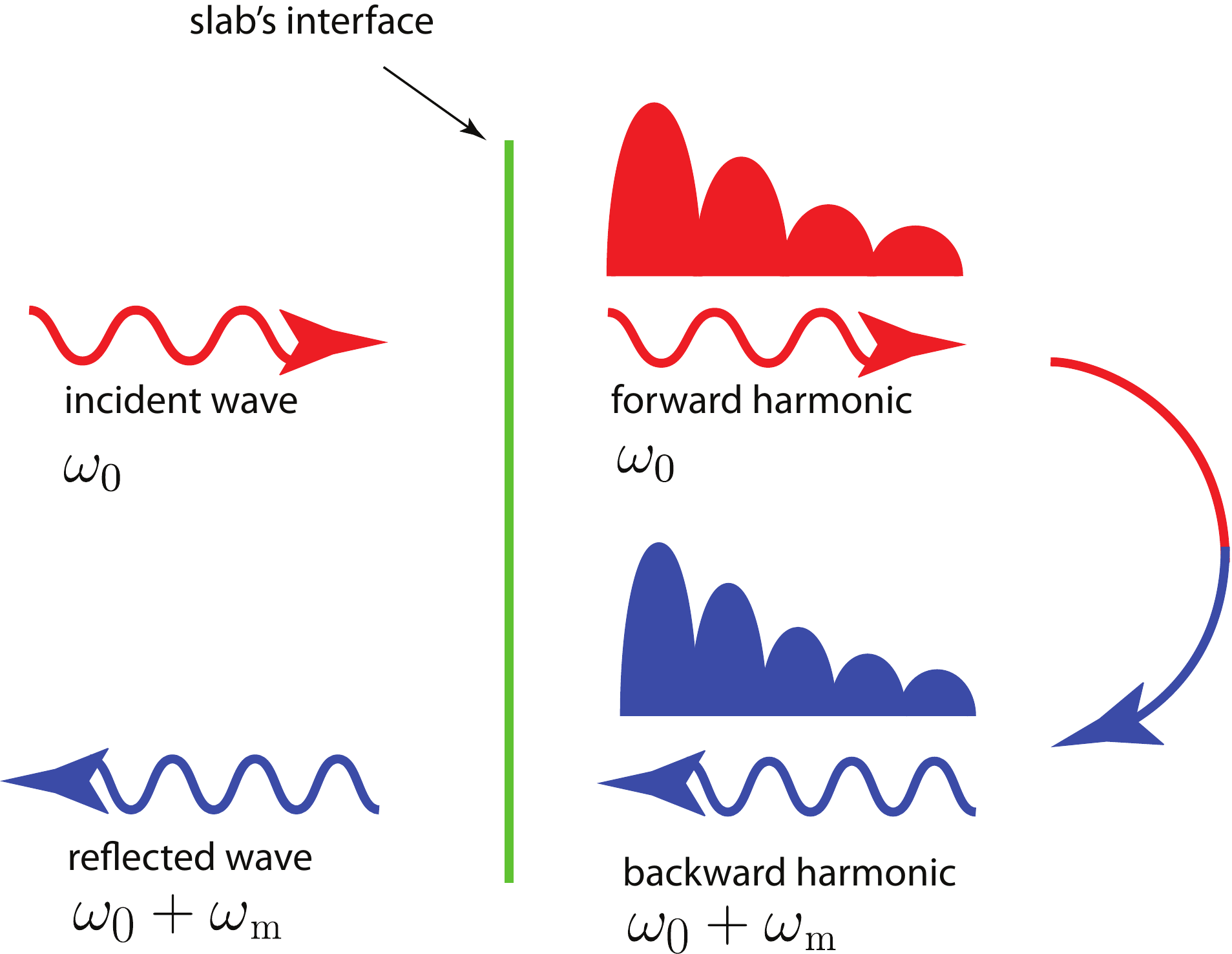}
\caption{Transformation of an incident forward propagating wave to an up-shifted backward propagating wave. An incident field with frequency $\omega_0$ impinges on the slab from the left side. The only 2 significant harmonics of the gap mode, at frequencies $\omega_0$ and $\omega_0+\omega_\text{m}$, are represented at the right side, where $\omega_\text{m}$ is the modulation frequency. The incident wave is fully matched to the evanescent harmonic at frequency $\omega_0$, therefore it excites this harmonic without back-reflection. The excited evanescent harmonic at frequency $\omega_0$ decays exponentially inside the slab, and its energy is transferred to the backward propagating harmonic at frequency $\omega_0+\omega_\text{m}$, whose energy increases exponentially as it reaches the slab interface. The backward propagating harmonic is fully matched to the incident region. It excites the reflected wave at frequency $\omega_0+\omega_\text{m}$ without any back-reflection inside the slab.}
\label{fig:redshift-demonstration}
\end{figure}

To see how the forward propagating wave is transformed into a backward propagating wave at an up-shifted frequency when the structure is excited from the left, it is instructive to inspect the electric and magnetic field profiles of the dominantly excited gap mode. The electric field profile for the gap mode [red index~0 in Fig.~\ref{fig:st-inf-disp-modes}] and the corresponding temporal frequency spectrum are plotted in Fig.~\ref{fig:E-profile-gapmode} and Fig.~\ref{fig:st-harmonics-gapmode}, respectively. This evanescent mode has two dominant frequency harmonics, one at $\omega_0$ and one at $\omega_0+\omega_\text{m}$, where $\omega_0$ is the incident frequency and $\omega_\text{m}$ is the modulation frequency. The remaining frequency harmonics are at least $50$~dB weaker, and may hence be safely ignored. The ratio of the magnetic to the electric field $\eta_0H/E$ for each harmonic, i.e. the corresponding effective refractive index, normalized to the refractive of the incident region $n_0=\sqrt{\epsilon_r}$, is plotted in Fig.~\ref{fig:gapmode-impedance}. The harmonic at the fundamental frequency ($\omega_0$) is forward propagating, $n_\omega>0$, and completely matched to the incident medium, i.e. $\eta_\omega=\eta_0=1/\sqrt{\epsilon_\text{r}}$. The harmonic at frequency $\omega_0+\omega_\text{m}$ has a negative effective refractive index $n_{\omega+\omega_\text{m}}=-n_0$. Therefore, this harmonic propagates backward, while being also fully matched to the incident medium. The blue-shift mechanism is schematically explained in Fig.~\ref{fig:redshift-demonstration}. As the incident wave and the forward harmonic are fully matched and have the same frequency, $\omega_0$, the incident wave excites this harmonic without any reflection at frequency $\omega_0$. The forward harmonic is evanescent and exponentially decays inside the slab. As it decays, it is converted to the backward propagating harmonic, which exponentially grows towards the interface with frequency $\omega_0+\omega_\text{m}$. This effect is clearly seen in the time domain simulation of the gap mode (see animation in supplemental material~\cite{supplementalMaterial}). The backward harmonic, which is also fully matched to the incident region, then excites the reflected wave at frequency $\omega_0+\omega_\text{m}$ when it hits the interface, without any back-reflection inside the slab.


\section{Comparison with Moving System}

It should be noted that asymmetric photonic bandgaps can also be produced in moving photonic crystals~\cite{wang2013opticaldiode}. Consider a moving photonic crystal slab with a gap at frequency $\omega_0$ in its reference frame. Assume that the crystal moves with constant velocity towards to the left, and that a wave impinges on it from the left with frequency $\omega_F$. In the reference frame of the moving photonic crystal, this incident wave is blue-shifted by the frequency amount $\Delta\omega$ corresponding to the relativistic Doppler effect, and it would be reflected if this blue-shifted frequency fell in the bandgap of the reference frame of the photonic crystal, or if $\omega_F+\Delta\omega=\omega_0$. In other words, for a static observer, the bandgap would appear red-shifted to the frequency $\omega_F=\omega_0-\Delta\omega$. Similarly, in the reference frame of the moving photonic crystal, a wave incident from the right with frequency $\omega_B$ would be perceived as red-shifted by the frequency amount $\Delta\omega$, and it would be reflected if $\omega_B-\Delta\omega=\omega_0$, i.e. to the frequency $\omega_B=\omega_0+\Delta\omega$ for the static observer the gap is blue shifted.  Thus, in the reference frame of the static observer, the photonic bandgaps are asymmetric. 

Similar to space-time modulated media, such asymmetry might be leveraged for the realization of nonreciprocal optical devices~\cite{wang2013opticaldiode}. However, despite similarities, moving media and space-time modulated systems have very distinct natures. A moving medium produces a drag effect (Fizeau drag). As a result, forward and backward harmonics appear to propagate with different group velocities in the reference frame of a static observer. In contrast, space-time modulated media do not alter the group velocities of the forward and backward harmonics, compared to the group velocities in the unmodulated medium. Moreover, in the case of a moving (isotropic) medium, the material parameters appear bianisotropic to a static observer due to the drag effect~\cite{plebanski1960electromagnetic, thompson2010completely, tretyakov2008generalized}. This complexity is eliminated by Lorentz transformation to the reference frame of the moving medium, where the medium becomes static and hence again isotropic. In contrast, space-time modulation does not alter the constitutive relations, i.e. an space-time modulated isotropic material would remain isotropic. However, there is generally no frame of reference that can transform a space-time modulated medium to a completely static medium.

\section{Isolation, Modulation and Bandwidth} \label{sec:discussion}

\begin{figure}[ht!]
\includegraphics[width=1.0\columnwidth]{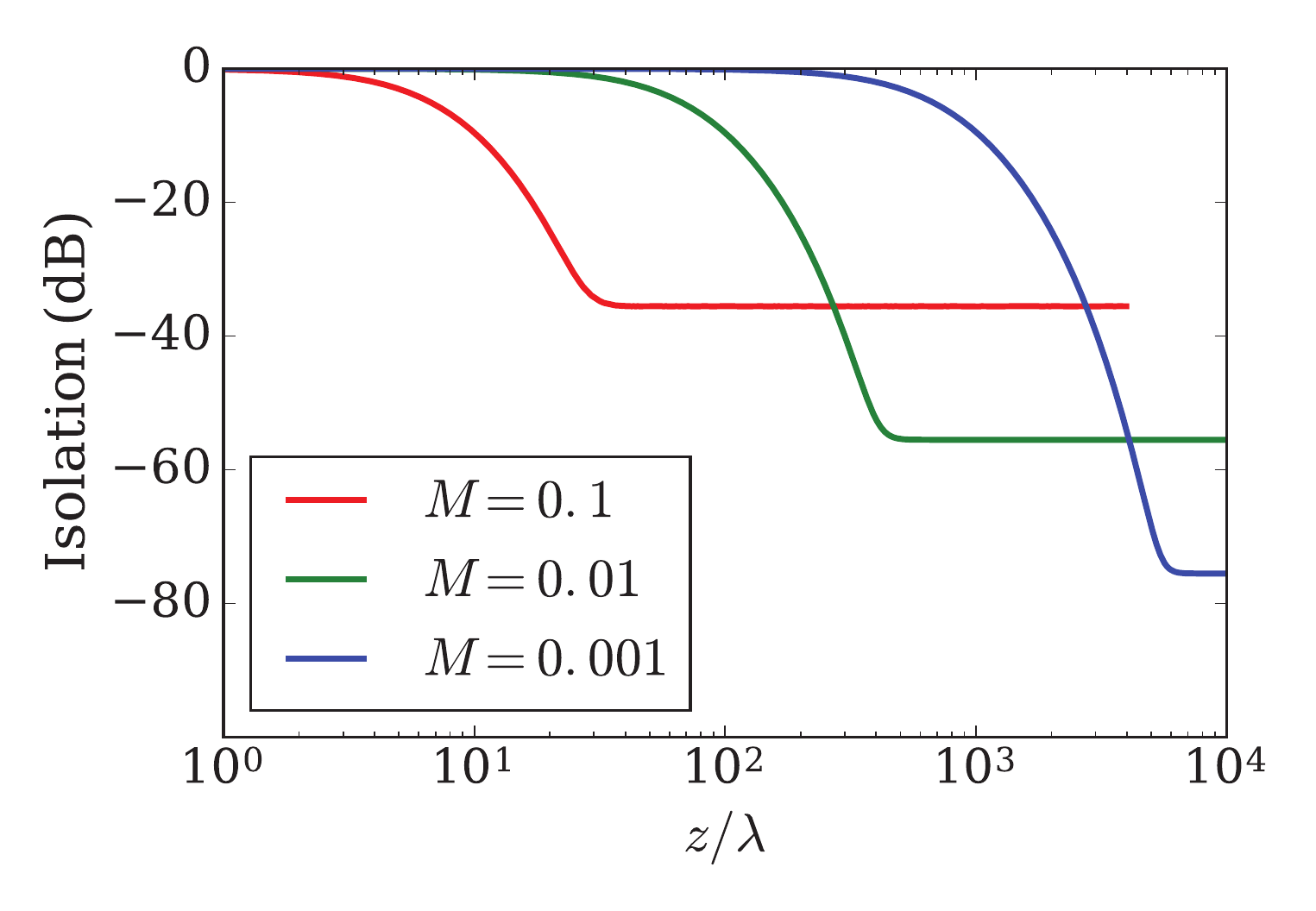}
\caption{Isolation versus modulated slab length for different modulation depths. For weaker modulations, longer space-time sections are required to achieve a specific amount of isolation. For each modulation depth, the isolation saturates at a specific level due to coupling to the undesired propagating modes, which act as a noise floor. For weaker modulations, coupling to undesired modes is weaker and higher isolation levels are achievable.}
\label{fig:isolation}
\end{figure}

The proposed spacetime system can achieve very high isolation levels even for extremely weak modulations. This is achieved by a sufficiently long space-time modulated region. Isolation levels for different modulation depths and slab lengths are plotted in Fig.~\ref{fig:isolation}. As the modulation becomes weaker, longer space-time sections are required to get the desired isolation levels. Note that for each modulation depth, isolation saturates at a specific level as the length of the space-time modulated section is increased. This saturation is caused by coupling to the undesirable modes of the space-time slab (red indices $-1, 1, -2, 2,\ldots$ in Fig.~\ref{fig:st-inf-disp-modes}), whose level represent an effective noise floor to the desired wave. For weaker modulations, coupling to the undesirable modes is weaker, and therefore higher isolation levels are achievable.

The required modulation frequency for creating the required asymmetric bandgaps is relatively low. It is proportional to the width of the bandgap, which is directly proportional to the modulation depth. For lower modulation levels, the bandgaps are narrower, and therefore it takes a smaller modulation frequency to misalign the forward and backward gaps. For very small modulation depths ($M\ll1$), the width of the first bandgap can be approximated as $\Delta\omega/\omega_0 = 2M/\left[\pi(1-M)\right]$~\cite{joannopoulos2011photonic}. Therefore, a modulation frequency in the order of $\omega_\text{m}=\Delta\omega = 2M/\left[\pi(1-M)\right]$ is sufficient to displace the bandgaps to as to achieve nonreciprocity. Decreasing the modulation depth, reduces the required modulation frequency as much as desired. For modulation depths smaller than $10^{-7}$, optical isolation may be achieved through ultrasound waves. However, the isolation bandwidth would be proportionally small, and the length of the device would be proportionally long. In such a case, the waveguide may be folded into a space-time modulated ring resonator for device footprint reduction~\cite{yu2009complete}.

\newpage

\begin{figure}[ht!]
\includegraphics[width=1.0\columnwidth]{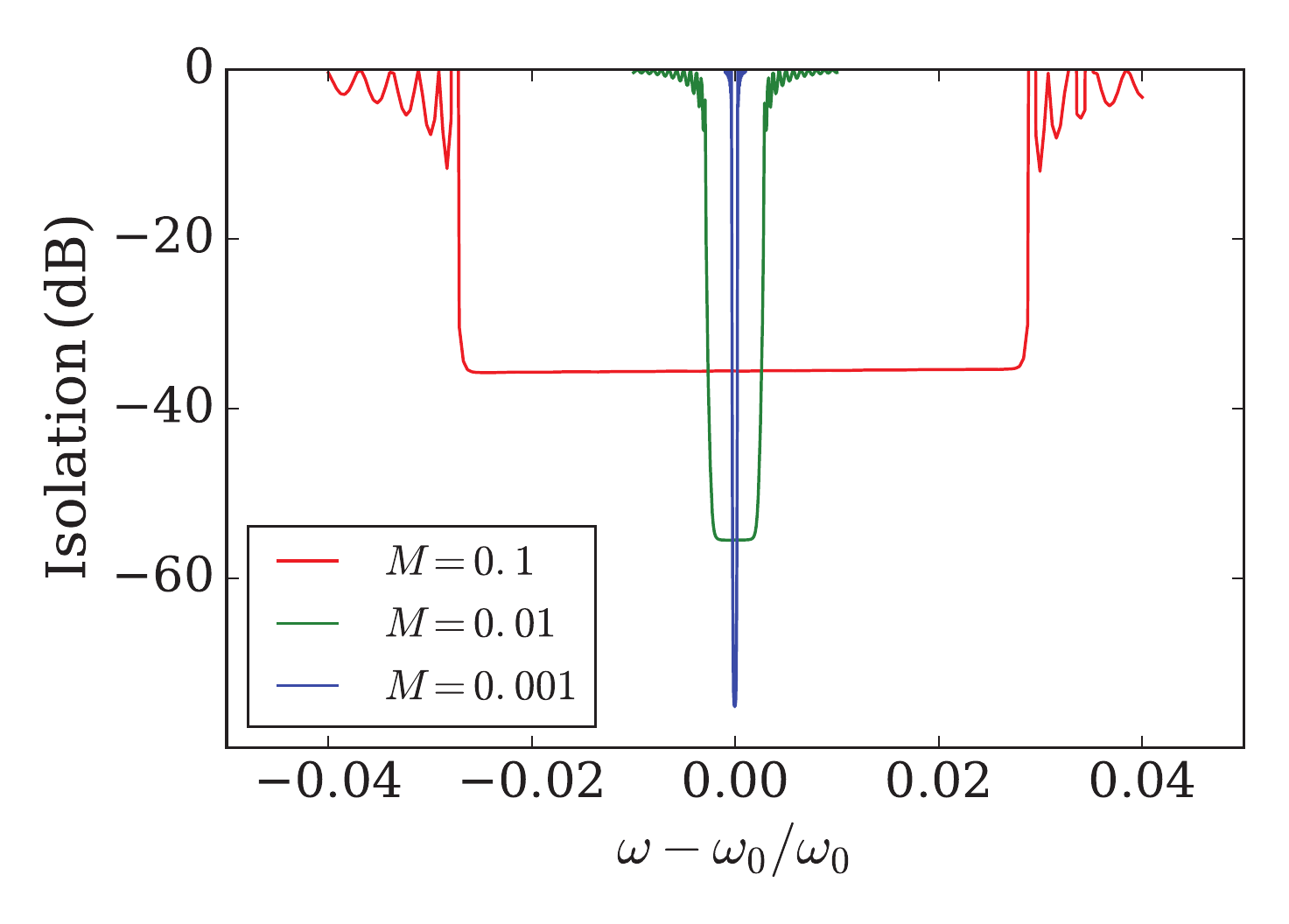}
\caption{Isolation versus frequency for different modulation depths and slab lengths corresponding to the knee points in Fig.~\ref{fig:isolation}. The bandwidth is directly proportional to the modulation depth.}
\label{fig:bandwidth}
\end{figure}

The nonreciprocity operation bandwidth is directly proportional to the width of the bandgap since nonreciprocity is produced by the bandgaps. The isolation versus frequency for different modulation depths is plotted Fig.~\ref{fig:bandwidth}. The lengths of the space-time slabs are chosen according to the saturation isolation knee points in Fig.~\ref{fig:isolation}. For modulation depths $M=0.1$, $M=0.01$ and $M=0.001$, the bandwidth is less than $5\%$, $1\%$ and $0.1\%$, respectively. A given bandwidth and isolation level may be achieved from an interplay between the modulation depth, and the length of the space-time modulated section.

\section{Up and Down Conversion Reflection Mixer}

\begin{figure}[ht!]
\includegraphics[width=1.0\columnwidth]{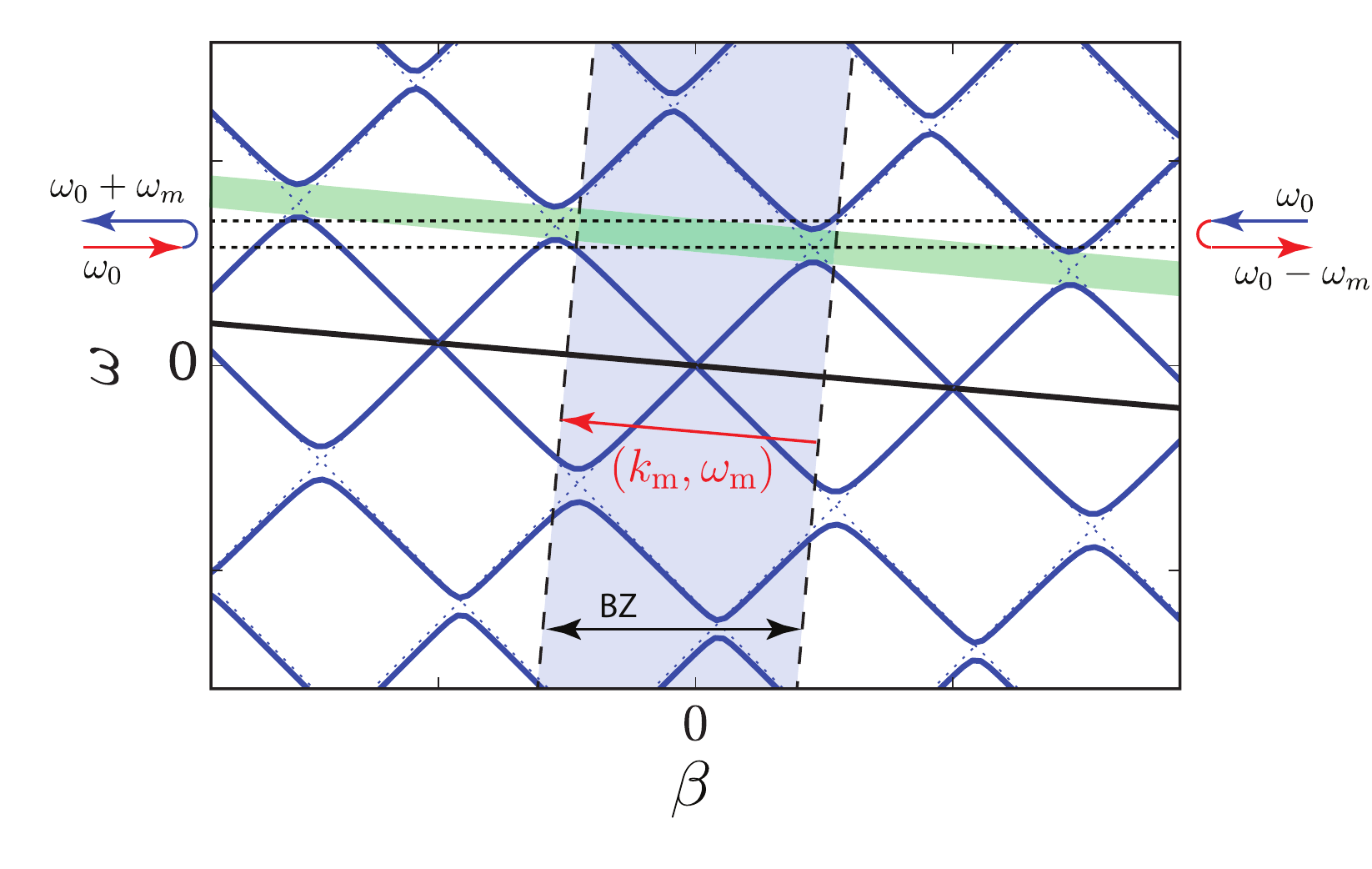}
\caption{Operation of the space-time modulated slab as a reflection-type mixer. When the structure is excited from the left at the down-tilted forward bandgap, the wave is fully reflected and blue-shifted. When the structure is excited from the right at the up-tilted backward bandgap, the wave is fully reflected and red-shifted. }
\label{fig:mixer}
\end{figure}

As the forward excited wave is fully reflected at a shifted frequency, the structure can also operate as a reflection-type optical mixer. Assuming modulation propagation to the left, the incident signal is up-shifted if the structure is excited from the left at the first bandgap, and down-shifted when the structure is excited from the right at the second bandgap, as depicted in Fig.~\ref{fig:mixer}. The amount of frequency shift is directly proportional to the modulation frequency. The mixing operation is almost perfect as the incident power is almost fully transferred to the desired up- or down-shifted frequency, without generating undesirable harmonics and inter-modulation products.

\section{Experimental demonstration} \label{sec:experiment}
\begin{figure}
\subfigure{ \label{Fig:isolator_circuit_schematic}
\includegraphics[width=\columnwidth]{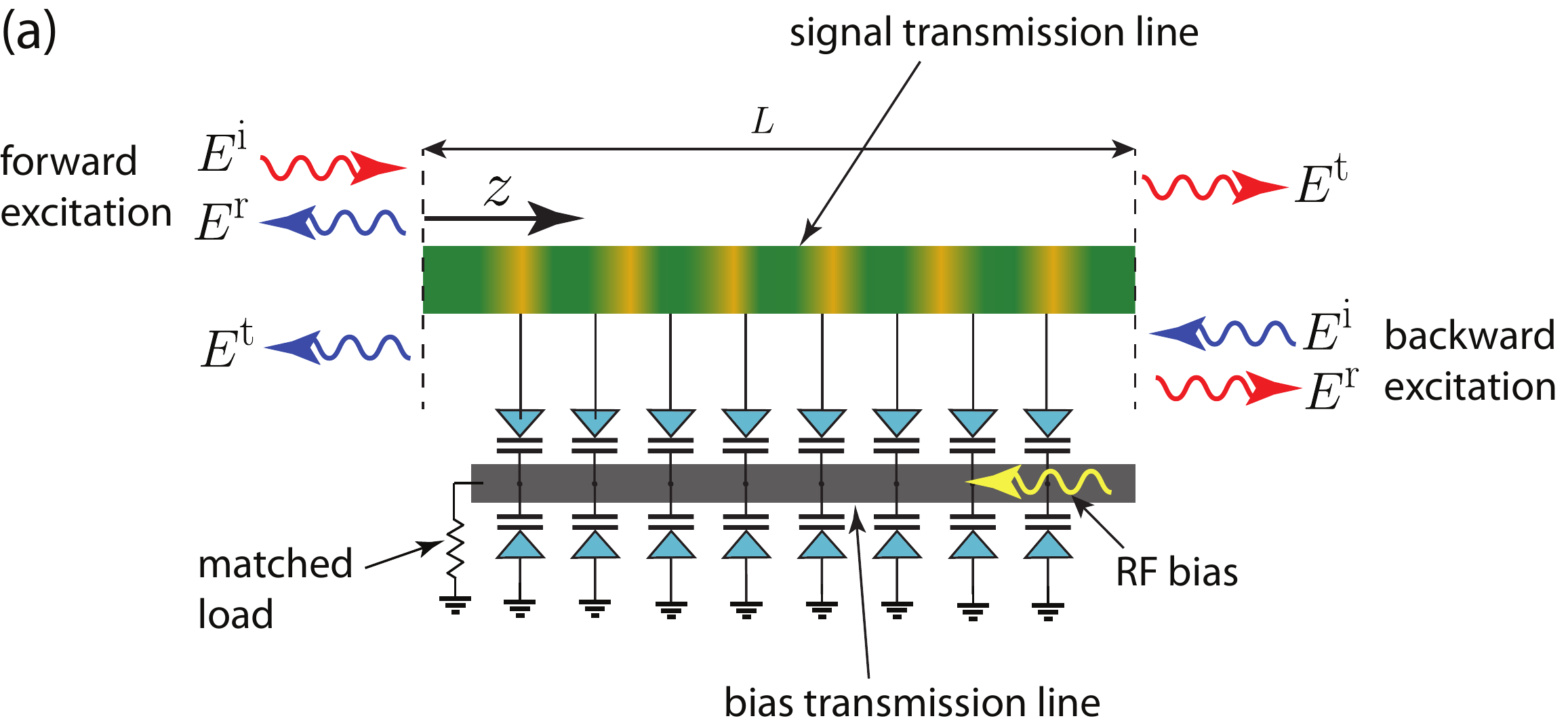}}
\subfigure{ \label{Fig:isol_photo}
\includegraphics[width=\columnwidth]{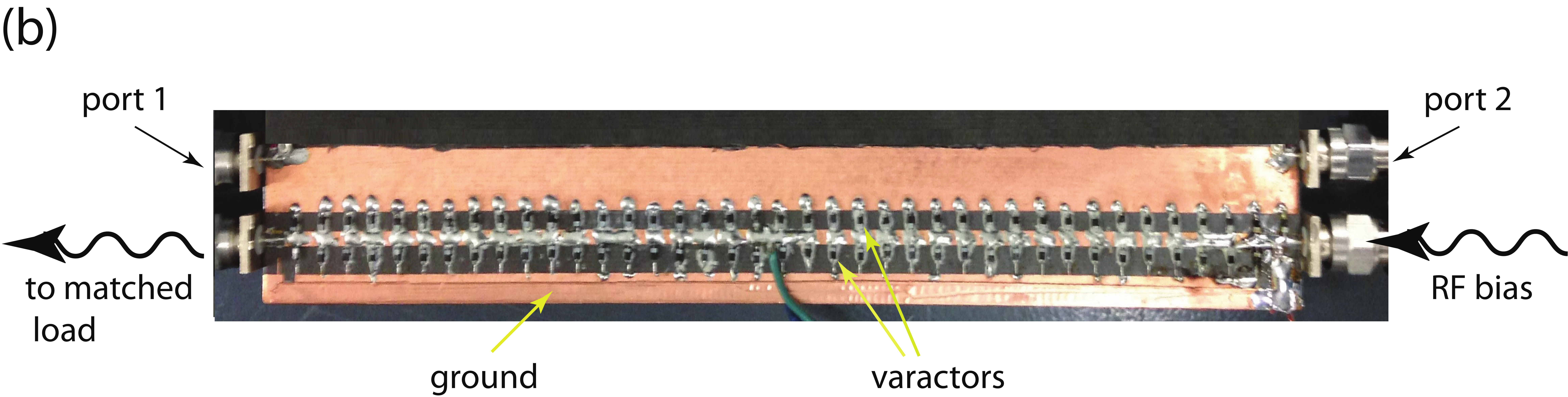}}
\caption{Experimental realization of the space-time varying system in the form of a space-time varying artificial microstrip transmission line. (a)~Schematic of the system, with distributed-capacitance varactors modulated by a radio wave emulating~\eqref{eq:cos-profile}. (b)~Photograph of the fabricated structure. The varactors were are the BB833 from Infineon Technologies, with capacitance ratio $C_\text{max}/C_\text{min} = 12$. The structure is $L=8$~inches long and is excited at $\omega_0=2\pi\times2.5$~GHz. The substrate is RT6010 from Rogers with permittivity~$10.2$, thickness~$h = 100$~mil and $\tan \delta = 0.0023$.}
\label{Fig:isolator_circuit}
\end{figure}

The space-time modulated system was realized at microwave frequencies in the form of a space-time varying artificial microstrip transmission line shown in Fig.~\ref{Fig:isolator_circuit}. In order to provide spatio-temporal control on the distributed capacitance of the transmission line, it is loaded with an array of sub-wavelengthly spaced shunt varactors. The bias line at the bottom provides a DC bias $V_\text{DC}$ plus a propagating RF bias,
\begin{equation}
V(z,t) = V_\text{DC} + V_\text{m}\cos(\omega_\text{m}t + k_\text{m}z)
\end{equation}

\noindent
to the varactors, where $\omega_\text{m}$ is the modulation frequency. The bias phase velocity $v_\text{m} = \omega_\text{m}/k_\text{m}$ is related to the bias line per-unit-length capacitance ($C_\text{av}$) and inductance ($L_\text{av}$) by $v_\text{m} = 1/\sqrt{L_\text{av}C_\text{av}}$. The varactors are reverse biased and act as voltage controlled capacitors. They thus add the space-time varying distributed capacitance

\begin{equation} \label{eq:st-capcitance}
C(z,t)= C_\text{av}+C_\text{m}~\cos(\omega_\text{m}t+k_\text{m}z)
\end{equation}

\noindent
to the signal transmission line. The structure in Fig.~\ref{Fig:isolator_circuit} therefore emulates a material with space-time varying permittivity~\eqref{eq:cos-profile}, with background permittivity $\epsilon_r \propto C_\text{av}$ and modulation depth $M=C_\text{m}/C_\text{av}$.

\begin{figure}
\includegraphics[width=0.95\columnwidth]{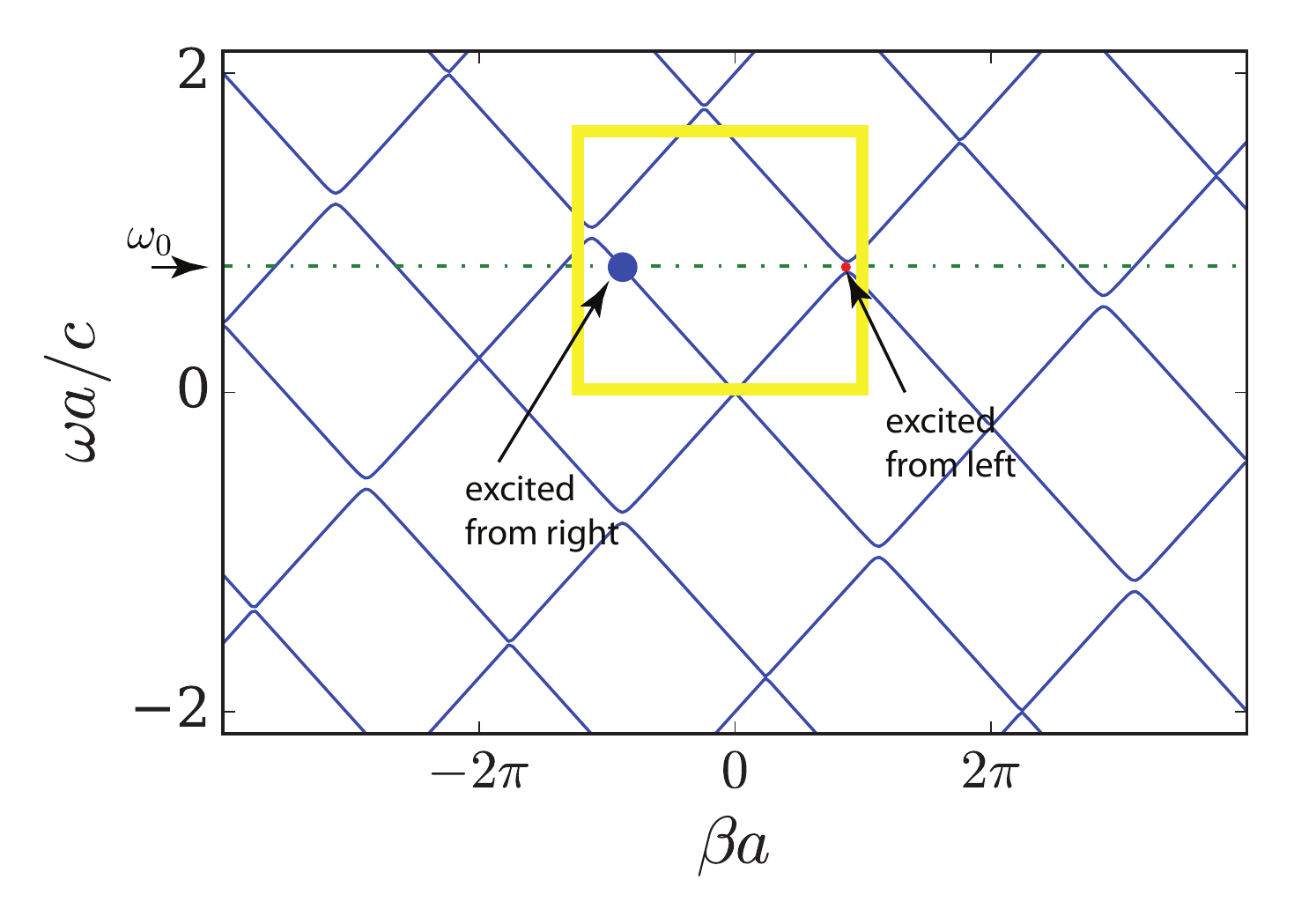}
\caption{Dispersion diagram corresponding to~\eqref{eq:st-capcitance} with parameters $\omega_\text{m}=2\pi\times0.675$~GHz, $k_\text{m}=415.79$~rad/m and $M=C_\text{m}/C_\text{av}=0.15$ in the structure of Fig.~\ref{Fig:isolator_circuit}. The yellow window corresponds to the asymmetric bandgap structure in Fig.~\ref{fig:concept_asymgap}. The red and blue dots represent the dominantly excited mode for forward and backward excitations, respectively.}
\label{Fig:dispesion_diag}
\end{figure}

\begin{figure}
\subfigure{\label{Fig:iso_res_forward}
\includegraphics[width=0.95\columnwidth]{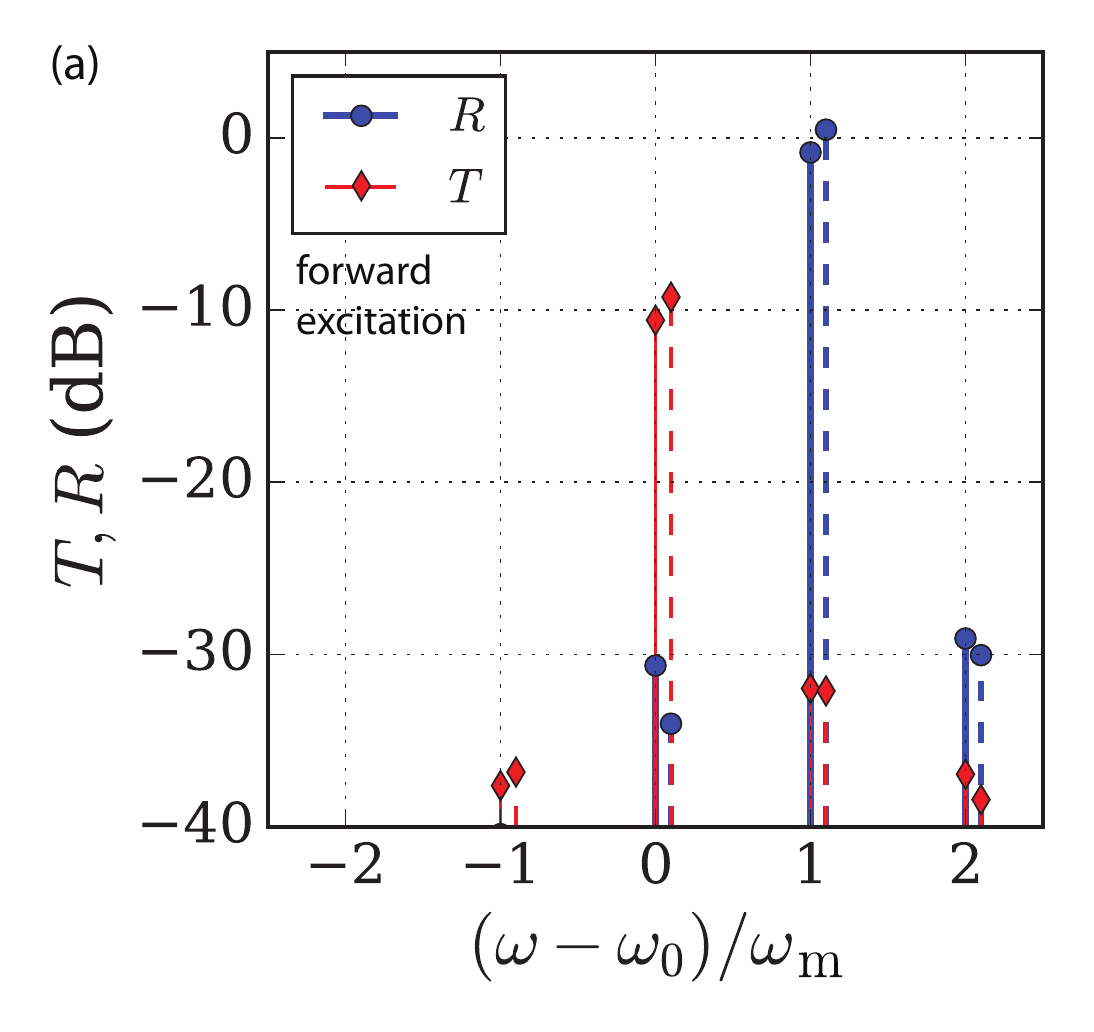}}
\subfigure{\label{Fig:iso_res_backward}
\includegraphics[width=0.95\columnwidth]{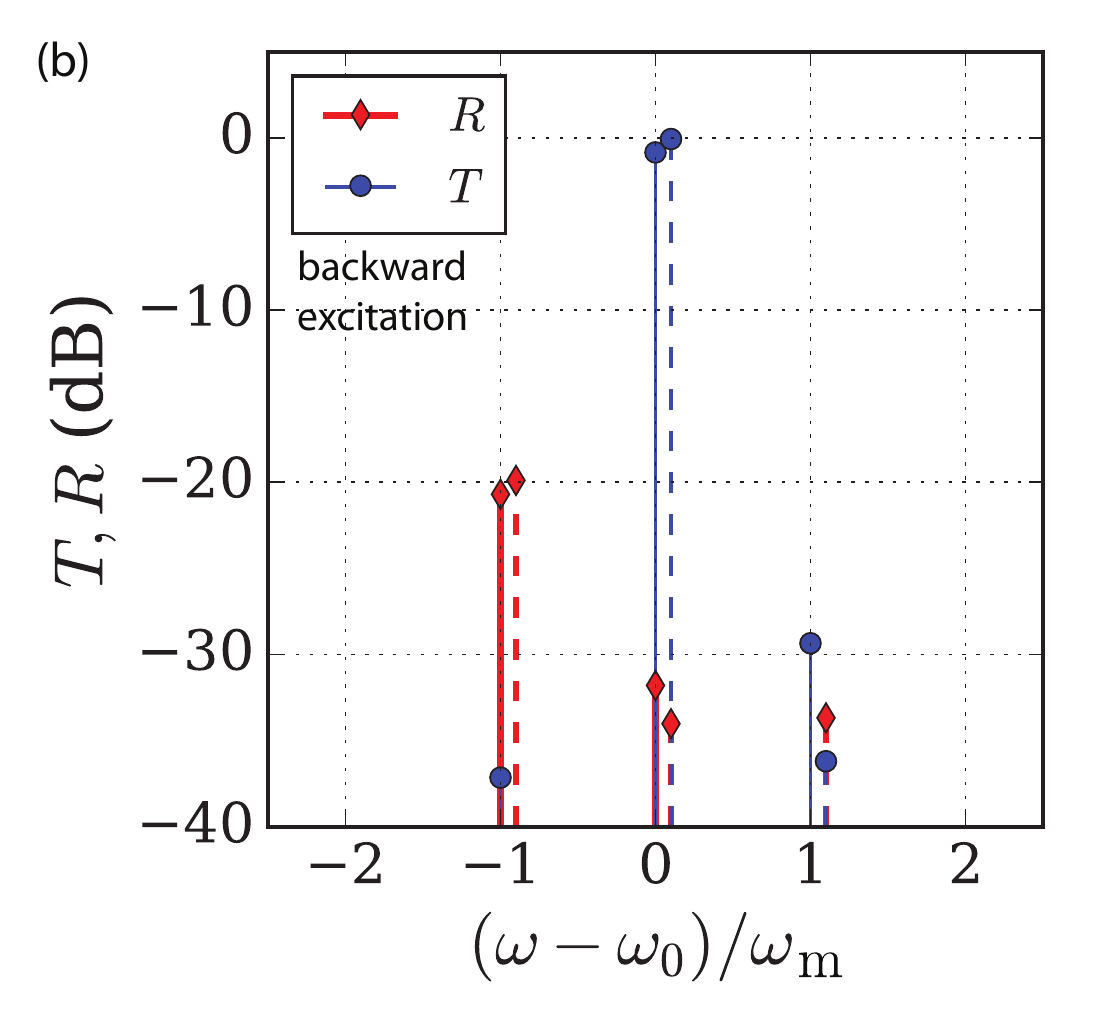}}
\caption{Experimental (solid lines) vs. theoretical (dashed lines) results for the isolator in Fig.~\ref{Fig:isolator_circuit} for the same parameters as in Fig.~\ref{Fig:dispesion_diag}. (a)~Forward excitation: the wave is almost fully reflected at the blue-shifted frequency $\omega_0+\omega_\text{m}=2\pi\times3.175$~GHz with a transmission level less than $-10$~dB. (b)~Backward excitation: the backward incident wave is fully transmitted at $\omega_0=2\pi\times2.5$~GHz. For clarity, the theoretical results are shifted by $0.1\omega_\text{m}$.}
\label{Fig:iso_res}
\end{figure}

Figure~\ref{Fig:isolator_circuit} shows a photograph of the space-time varying microstrip line. The modulation circuit is comprised of 39 unit cells of antiparallel varactors, uniformly distributed along the microstrip line, with the subwavelength period $p=5$~mm, corresponding to $p/\lambda_\text{m}\approx 1/19$. Therefore, effectively, the structure represents a medium with the continuous permittivity~\eqref{eq:cos-profile}. The corresponding dispersion curves are plotted in Fig.~\ref{Fig:dispesion_diag}, where the horizontal line represents the excitation frequency. The incident frequency is chosen to excite the evanescent mode marked by the red dot in the forward direction and the propagating mode marked by the blue point in the backward direction. The corresponding length at this frequency is $L=6 \lambda_0$.

The scattering parameters are plotted in Figs.~\ref{Fig:iso_res_forward} and~\ref{Fig:iso_res_backward} for forward and backward excitations, respectively. The evanescent mode decays by 10.5~dB before reaching the end of the structure, corresponding to -10.5~dB transmission in Fig.~\ref{Fig:iso_res_forward}. The rest of the power is reflected at the up-shifted frequency $\omega_0+\omega_\text{m}=2\pi\times 3.175$~GHz. In the backward direction the incident wave is almost fully transmitted. Therefore, the isolation level is 10.5~dB. Higher isolation levels may be achieved by increasing the length of the structure. The small discrepancy between theory and experiment are attributed to the metallic and dielectric losses in the experiment that have not been accounted for in the theory.
\vspace{2cm}

\section{Conclusions} \label{sec:conclusion}
Space-time modulation has been introduced as a technique to tilt the band structure of photonic crystals, resulting in asymmetrically-aligned photonic bandgaps for opposite directions of propagation. Such space-time modulated slabs have been excited at the frequency corresponding to a photonic bandgap, exciting the evanescent bandgap mode in the forward direction while exciting a propagating mode in the opposite direction. Using a full-wave modal analysis, it has been shown that in the forward direction all the energy is reflected at a Doppler shifted frequency. In the opposite direction, the incident wave is fully transferred to the other end of the space-time modulated slab by strongly coupling to one of its propagating modes, hence realizing an optical isolator and a reflection-type mixer.

\bibliography{ReferenceList2_abbr}

\end{document}